# Programmable Intrinsic Circularly Polarized Emission

Yujie Jiao[1,2†], Zhenqin Li[2†], Yide Chang[3†], Yongsen He[4,7], Xiaoyu Sun[5], Wei Lyu[6], Jiayu Ding[2], Zhaolin Zhong[2], Puxin Yang[2], Chi Chen[3], Bangjie Song[2], Min Qiu[5,6*], Yanming Wang[7*], and Siying Peng[2*]

[1]School of Materials Science and Engineering, Zhejiang University, Hangzhou, Zhejiang 310027, China.
[2]Department of Materials Science and Engineering, School of Engineering, Westlake University, Hangzhou, Zhejiang 310030, China.
[3]Global College, Shanghai Jiao Tong University, Shanghai 200240, China.
[4]School of Mechanical Engineering, Shanghai Jiao Tong University, Shanghai 200240, China.
[5]Zhejiang Key Laboratory of 3D Micro/Nano Fabrication and Characterization, Department of Electronic and Information Engineering, School of Engineering, Westlake University, Hangzhou, Zhejiang 310030, China.
[6]Westlake Institute for Optoelectronics, Fuyang, Hangzhou, Zhejiang 311400, China.
[7]Global Institute of Future Technology, Shanghai Jiao Tong University, Shanghai 200240, China.

†These authors contributed equally to this work.

Corresponding authors: pengsiying@westlake.edu.cn; yanming.wang@sjtu.edu.cn; qiu_lab@westlake.edu.cn.

## Abstract

Circularly polarized luminescence (CPL) is central to chiral photonics, yet programming circularly polarized emission at the nanoscale remains challenging. Here, we program intrinsic CPL at its microscopic origin in laser-written all-inorganic perovskite nanocrystals embedded in glass. High-resolution transmission electron microscopy reveals a core-shell-like variation in interplanar spacing associated with intrinsic CPL, consistent with torsional lattice distortion. The torsional lattice distortion breaks inversion symmetry, while density functional theory calculations show that it lifts the spin degeneracy of the band-edge electronic states. Power-dependent measurements further reveal a transition from birefringence-mediated circular polarization to intrinsic CPL, accompanied by the emergence of a distinct core-shell-like lattice distortion in the nanocrystals. By tuning the incident linear polarization angle and focal depth, we deterministically control both the handedness and magnitude of the intrinsic CPL, with $|g_{lum}|$ of approximately $4 \times 10^{-3}$. These results show that programmable intrinsic CPL originates from the structural and electronic properties of the emitting nanocrystals, enabling circularly polarized emission to be controlled at its microscopic origin and spatially encoded within a monolithic material.

## Introduction

Spatial inversion symmetry breaking plays a fundamental role in structural chirality and the resulting chiroptical phenomena, including circularly polarized luminescence (CPL).[1] CPL is of interest for applications in spin-based optoelectronics and quantum information processing.[2-4] CPL commonly relies on chirality imparted by chiral molecules.[5-10] While effective, these approaches encode chirality through the incorporated chiral components, making spatial programming of emission handedness within a single monolithic material challenging. Recent studies have shown that chirality can also emerge in initially achiral materials through mechanical or structural deformation, as demonstrated by the piezochiral effect,[11] moiré superlattices,[12-15] and crystals containing screw dislocations.[16] However, whether intrinsic CPL can be generated and programmed directly within nanoscale emitters remains largely unexplored.

Here, we program intrinsic CPL at its microscopic origin in all-inorganic $CsPbBr_3$ perovskite nanocrystals embedded in glass. We utilize direct laser writing to crystallize the nanocrystals within a glass matrix,[17-19] where they exhibit intrinsic CPL. High-resolution transmission electron microscopy reveals a core-shell-like variation in interplanar spacing, consistent with lattice distortion under mechanical confinement. We describe this structural variation using a torsional-distortion model. The torsional lattice distortion breaks inversion symmetry, and density functional theory calculations show that it lifts the spin degeneracy of the band-edge electronic states.

To further examine the origin of the intrinsic CPL, we evaluate the chiroptical responses as a function of the irradiation power densities in direct laser writing, revealing a transition from birefringence-mediated circular polarization to intrinsic CPL. Furthermore, by systematically varying the linear polarization angle and the focal depth of the writing laser, we deterministically program the handedness and the magnitude of the intrinsic CPL. Thermo-mechanical simulations further reveal polarization-dependent changes in the stress distribution and signed shear imbalance within the surrounding glass, suggesting a mechanical origin of the programmable CPL response. Based on these programmable chiroptical properties, we fabricate a multidimensional optical information-encoding system. Together, these results show that circularly polarized emission can be programmed at its microscopic origin through the structural and electronic properties of the emitting nanocrystals, while its handedness can be spatially defined within a monolithic material, providing a solid-state platform for spatially integrated CPL emitters and multichannel optical information technologies.[20]

## Results and Discussion

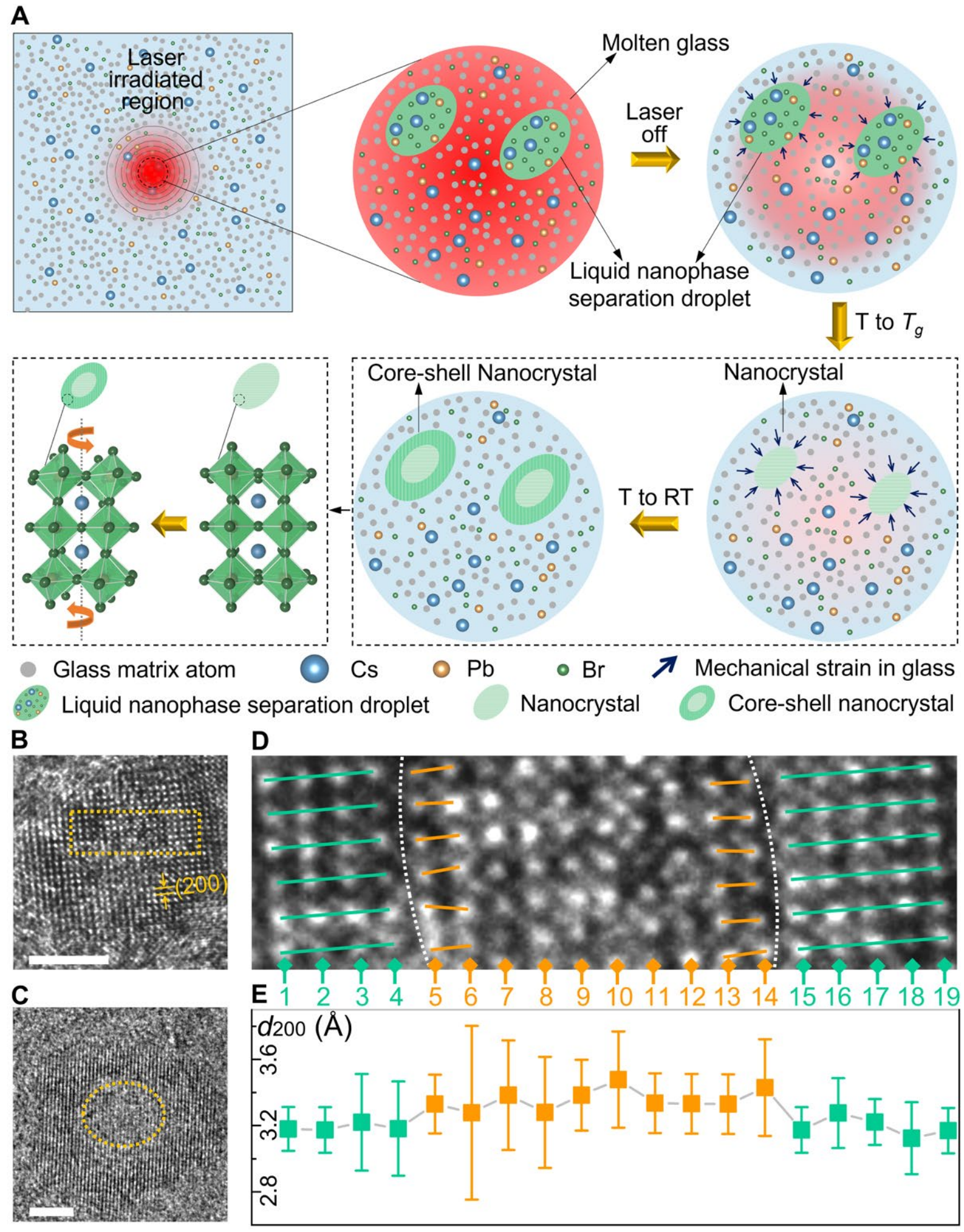

**Figure 1.** Formation of metastable core-shell-like nanocrystals in glass. (A) Schematic illustration of laser-induced crystallization. Multiphoton absorption generates localized temperatures that exceed the glass liquidus, producing liquid nanophase separation. Rapid quenching freezes an anisotropic residual stress field within the rigid glass matrix, resulting in the formation of metastable core-shell-like nanocrystals. $T_g$, glass transition temperature. RT, room temperature. Transmission electron microscopy images of core-shell-like nanocrystals in glass synthesized by direct laser writing at a power density of 63.4 TW/cm$^2$: (B) before and (C) after heat treatment at 350 °C for 10 h. Scale bar, 4 nm. (D) Enlarged image of the nanocrystal in (B) indicated by the yellow dashed box, with the core-shell-like boundary marked by white dashed lines. Green and orange lines indicate the difference in interplanar spacing between the core and shell-like regions. (E) Averaged interplanar spacings across the core-shell-like boundary.

To generate perovskite nanocrystals under mechanical confinement, we employed ultrafast laser writing to create a localized nonequilibrium thermal and stress environment for the crystallization of all-inorganic $CsPbBr_3$ nanocrystals within a rigid glass matrix (Figure 1A). Focused linearly polarized laser pulses (5 ps, 1030 nm) at a power density of 63.4 TW/cm$^2$ trigger nonlinear multiphoton absorption, producing transient temperatures above the glass liquidus under highly confined conditions. This localized excitation induces liquid nanophase separation, followed by rapid cooling that retains an anisotropic residual stress field within the laser-modified glass. Subsequent thermal annealing (350 °C, 10 h), performed below the glass transition temperature ($T_g$),[18] provides thermal activation for the precipitation and growth of $CsPbBr_3$ nanocrystals within the laser-written regions. These regions exhibit bright green photoluminescence under ultraviolet excitation (Figure S1).

The rigid glass matrix imposes mechanical confinement during nanocrystal formation, giving rise to a metastable core-shell-like structural state. We propose that rapid cooling constrains the outer region of the forming nanocrystal while continued lattice contraction generates a nonuniform strain distribution within the confined crystal (Figure 1A). High-resolution transmission electron microscopy (HRTEM) performed immediately after laser irradiation reveals a distinct core-shell-like variation in lattice structure (Figure 1B), with additional examples shown in Figure S2. The enlarged HRTEM image in Figure 1D reveals a central region separated from the surrounding shell-like region by discernible structural boundary. Quantitative analysis of the lattice fringes shows a systematic difference in the apparent interplanar spacing across these interfaces, with the spacing in the core region being 4.9-5.2% larger than that in the surrounding shell-like region (Figure 1E).

The observed difference in interplanar spacing is consistent with a torsional lattice distortion of the mechanically confined nanocrystal. During rapid cooling, constrained lattice contraction can generate a nonuniform strain state because relaxation of the nanocrystal is restricted by the surrounding glass. Under such confinement, part of the lattice mismatch may be accommodated through torsional lattice distortion of the shell-like region. Such a deformation would reduce the apparent interplanar spacing in the two-dimensional HRTEM projection and could therefore account for the measured difference between the core and shell-like regions. Within this interpretation, the observed core-shell-like structural heterogeneity represents a metastable distorted lattice state retained by mechanical confinement within the glass matrix. Additional evidence of strain accommodation is provided by crystallographic twinning. Individual nanocrystals exhibit coherent twin boundaries, including a (102) twin plane identified from the corresponding fast Fourier transform (FFT) analysis (Figure S3).

Thermal annealing at 350 °C for 10 h subsequently promotes further nanocrystal growth, as observed by HRTEM (Figure 1C). At this temperature, short-range ionic migration enables crystal growth, while the surrounding glass remains below $T_g$ and continues to provide mechanical confinement. Because large-scale viscous relaxation of the glass matrix is suppressed under these conditions, nanocrystal growth proceeds within the pre-existing anisotropic strain environment generated by laser processing. The structural heterogeneity established during the initial laser-induced crystallization is therefore retained as the nanocrystals grow during subsequent annealing.

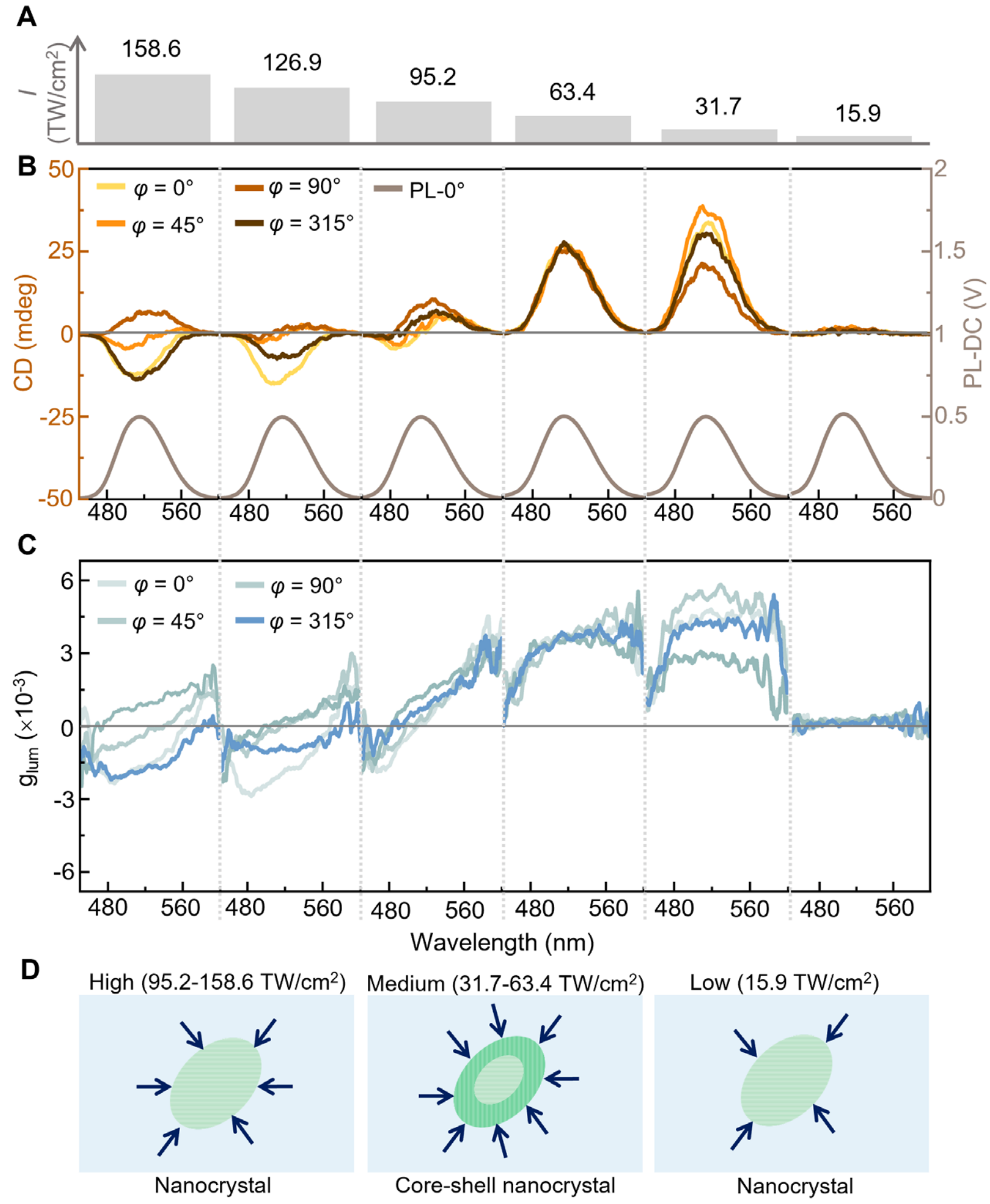

**Figure 2.** Intrinsic circularly polarized emission from core-shell-like nanocrystals. (A) Laser power densities used for sample fabrication. (B) Circularly polarized luminescence (CPL) spectra of nanocrystals synthesized at corresponding laser power densities (158.6, 126.9, 95.2, 63.4, 31.7, and 15.9 TW/cm$^2$), at different azimuthal angles, and their photoluminescence spectra, respectively. (C) The photoluminescence dissymmetry factor ($g_{lum}$) of samples synthesized at corresponding laser power densities and azimuthal angles. All samples were prepared with subsequent heat treatment at 350 °C for 10 h. (D) Schematic illustration of the laser-power-dependent structural evolution of the core-shell-like nanocrystal.

To determine the origin of the chiroptical response, we systematically measured the circularly polarized luminescence (CPL) of laser-written samples prepared over a broad range of irradiation power densities (15.9 to 158.6 TW/cm$^2$), followed by thermal annealing (350 °C for 10 h). CPL measurements were performed using a dedicated CPL spectrometer (JASCO CPL-300) with an unfocused excitation beam. To distinguish intrinsic CPL from circular polarization associated with linear optical anisotropies, CPL spectra were measured while rotating the sample about the optical axis at azimuthal angles ($\varphi$) of 0°, 45°, 90°, and 315° (Figures S4 and S5).

As shown in Figures 2B and 2C, three distinct chiroptical behaviors are observed across different ranges of laser irradiation power density. At high irradiation power densities (≥ 95.2

TW/cm$^2$), both the sign and magnitude of the CPL signal vary strongly with sample rotation. In particular, the azimuthal-angle-dependent sign reversal is characteristic of polarization conversion associated with linear optical anisotropy.[21,22] Previous polarimetric studies of laser-written glass have shown that nanograting-induced form birefringence and stress-induced birefringence with misaligned optical axes can give rise to extrinsic circularly polarized optical responses.[23] Such effects can contribute to the measured circular polarization in laser-modified glass and therefore need to be distinguished from intrinsic CPL of the emitters. In our samples, high-power density irradiation also produces pronounced laser-modified microstructures in the glass, consistent with an increased contribution from form birefringence to the measured polarization response. A different behavior is observed at intermediate irradiation power densities of 31.7 and 63.4 TW/cm$^2$. As shown in Figures 2B and 2C, the CPL signal retains the same handedness at all azimuthal angles. The persistence of the CPL sign upon azimuthal rotation distinguishes this response from the sign-changing polarization artifacts observed in the high-power regime and supports an intrinsic CPL from the emitting nanocrystals. The 63.4 TW/cm$^2$ condition yields a more uniform response and was therefore used for the subsequent measurements. Stokes polarimetry gives a degree of circular polarization (DOCP) of 0.008 and a degree of linear polarization (DOLP) of 0.1949 (Table S1). An independent analyzer-rotation measurement gives a DOLP of 0.1976 (Figure S6), in close agreement with the linear-polarization component obtained from the Stokes analysis.

At the lowest power density of 15.9 TW/cm$^2$, the CPL signal approaches the experimental baseline, indicating no detectable intrinsic CPL under these irradiation conditions. Photoluminescence spectra measured across intermediate- and high-power density regimes show similar emission peak positions and spectral widths, although the emission intensity varies with laser power density (Figure S7). The distinct CPL behaviors observed across the three regimes therefore cannot be accounted for by corresponding shifts or broadening of the photoluminescence spectra.

Control experiments were further performed to determine the processing conditions required for the intrinsic CPL response. Samples subjected to laser irradiation without subsequent thermal annealing exhibit only weak angle-dependent baseline fluctuations (Figure S8). Conversely, $CsPbBr_3$ nanocrystals produced by uniform thermal annealing at 455 °C without prior laser writing show no measurable CPL (Figure S9). These results indicate that the intrinsic CPL response requires the combination of localized laser processing and subsequent nanocrystal growth within the laser-modified glass.

The irradiation power density-dependent measurements therefore define three regimes: a high-power density regime dominated by rotation-dependent polarization responses associated with birefringence, an intermediate-power density regime exhibiting rotation-invariant CPL handedness, and a low-power density regime in which the CPL response remains near the experimental baseline (Figure 2D).

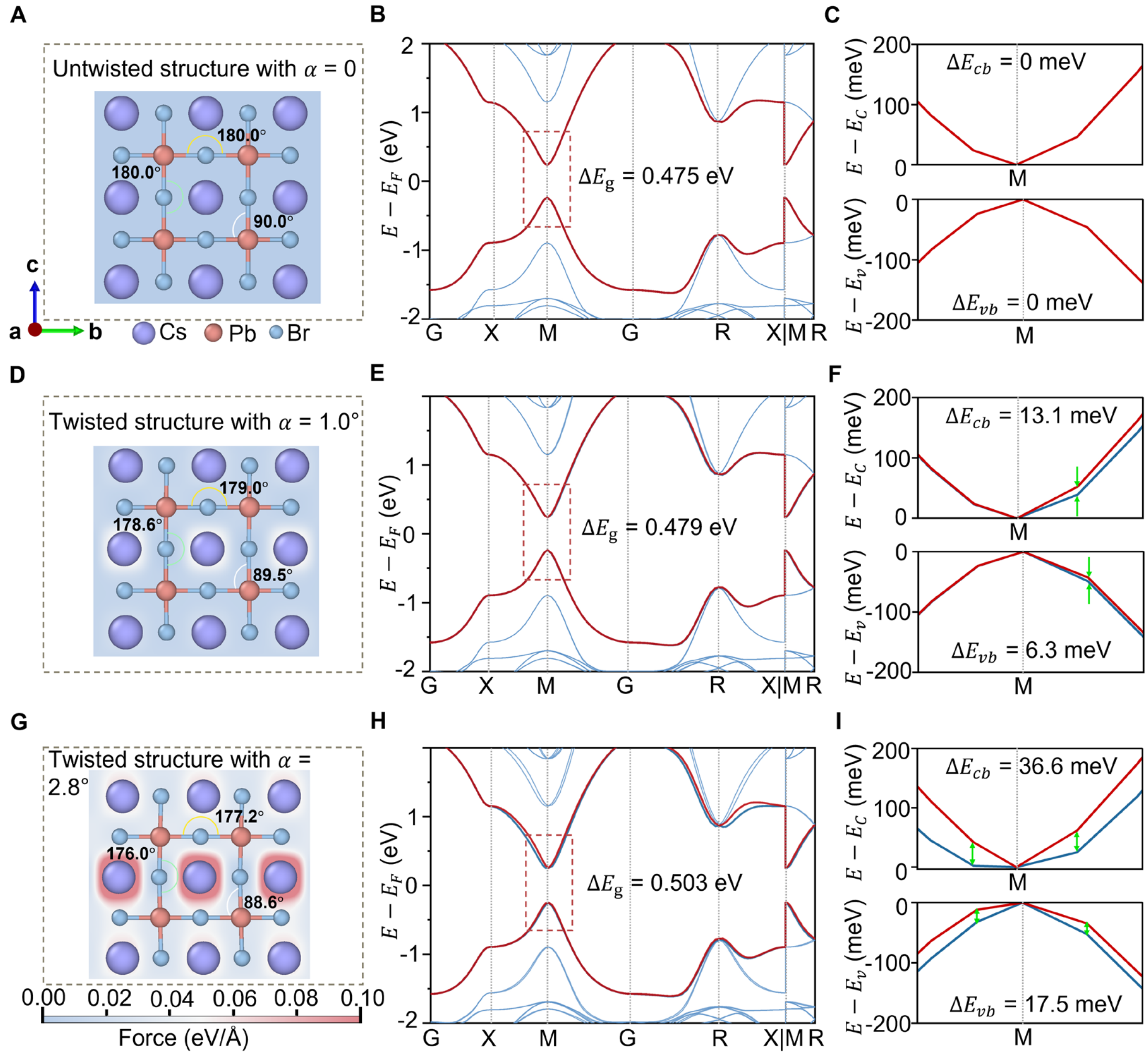

**Figure 3.** Density functional theory simulations for $CsPbBr_3$ with torsional lattice distortion. Atomistic structures of (A) untwisted $CsPbBr_3$ and twisted $CsPbBr_3$ with maximum rotation angles of (D) 1.0° and (G) 2.8°. The heat map shows the magnitude of the force on each atom using a color scale from 0 to 0.10 eV/Å; values above this range are saturated. The schematic on the right shows the layer-resolved rotation angles (such as 0°, 1.4°, 2.8°, 1.4° and 0° in (G)). The corresponding band structures are shown in (B), (E) and (H), respectively. Magnified conduction- and valence-band-edge dispersions near the M point, showing the torsion-dependent band splitting. The dashed boxes in (B), (E), and (H) indicate the band-edge regions enlarged in (C), (F), and (I), respectively.

To examine the origin of this non-monotonic power dependence, finite-element simulations (COMSOL) were used to model the evolution of the local stress field during laser heating and subsequent cooling. The temperature-dependent mechanical relaxation of the glass was described phenomenologically using the Williams-Landel-Ferry relation (Note S1, Tables S2 and S3, and Figure S10).[24] The calculated residual stress exhibits a non-monotonic dependence on laser power densities. At intermediate irradiation power densities, rapid cooling limits stress relaxation and allows a substantial residual stress field to remain within the laser-modified region. At higher power densities, the larger temperature rise produces stronger softening of the surrounding glass, increasing viscous relaxation before cooling is complete and thereby reducing the retained residual stress. Consistent with this interpretation, nanocrystals produced under high-power density irradiation do not exhibit the pronounced core-shell-like variation in interplanar spacing observed under the intermediate-power density condition, either before or after thermal annealing (Figure S11). Together, the simulations and structural observations support a processing window in which residual stress is preferentially retained during cooling, coinciding with the irradiation regime in which rotation-invariant intrinsic CPL is observed.

To examine how torsional lattice distortion modifies the electronic structure of $CsPbBr_3$, we performed density functional theory (DFT) calculations for pristine and distorted cubic structures. Figures 3A-3C display the atomic configuration, the electronic band structure, and magnified band-edge dispersion of the untwisted $CsPbBr_3$, respectively. In its pristine state, the cubic $CsPbBr_3$ exhibits a direct bandgap of 0.475 eV at the M point. Due to the presence of inversion and time-reversal symmetry in the lattice, the SOC-included bands remain spin-degenerate, with the $\Delta E_{cb}$ and $\Delta E_{vb}$ both equal to 0 meV (Figure 3C). Upon introducing a maximum torsional angle of 1.0°, the bandgap increases slightly to 0.479 eV, while the conduction- and valence-band splittings reach 13.1 and 6.3 meV. Further increasing the maximum rotation angle to 2.8°increases the bandgap to 0.503 eV and enhances the corresponding splittings to 36.6 meV and 17.5 meV, indicating a systematic electronic-structure response to the magnitude of the imposed torsional lattice distortion.

To construct a distortion model representative of the experimentally observed lattice distortion, a gradient torsional lattice distortion was applied along the c-axis of the supercell. Specifically, based on the apparent interplanar-spacing mismatch of approximately 4.9% observed in the HRTEM analysis, a maximum rotation angle of 2.8° was introduced at the center of the supercell (Details in Figure S12). The torsional angle progressively decreased to 1.4° in the adjacent layers, while the outermost layers were rigidly fixed at 0° to maintain bulk-like boundary conditions and preserve structural continuity. The lattice constants of the distorted model were relaxed while the atomic positions were held fixed. This gradient configuration was used to represent the spatially nonuniform lattice distortion observed in the mechanically confined nanocrystals (Note S2). The corresponding structural distortion parameters are summarized in Table S4. For comparison, a second gradient torsional model with a reduced maximum rotation angle of 1.0° was constructed using the same spatial profile.

This torsion-induced band splitting is consistent with Rashba-type spin splitting.[9,25] The gradient torsional lattice distortion breaks the spatial inversion symmetry of the lattice, creating an asymmetric crystal potential. In the presence of strong intrinsic spin-orbit coupling (SOC) associated with the heavy Pb atoms, this inversion symmetry breaking lifts the spin degeneracy, splitting the band-edge electronic states into spin-polarized branches. Such spin-polarized band-edge states are expected to modify the optical selection rules and can give rise to unequal transition probabilities for left- and right-circularly polarized emission. The more pronounced band splitting in the 2.8° model further suggests that the spin-dependent electronic structure is sensitive to the magnitude of the torsional lattice distortion, which may contribute to the experimentally observed variation in the chiroptical response. No electronic band splitting was observed for the uniaxial tensile or compressive strain, biaxial tensile or compressive strain, shear strain, or combined normal-strain condition, with tension along the $a$ and $c$ axes and compression along the $b$ axis, under the conditions examined here (Figures S13-S18). These control calculations indicate that the band splitting observed in the torsionally distorted models is not reproduced by the conventional strain configurations examined here.

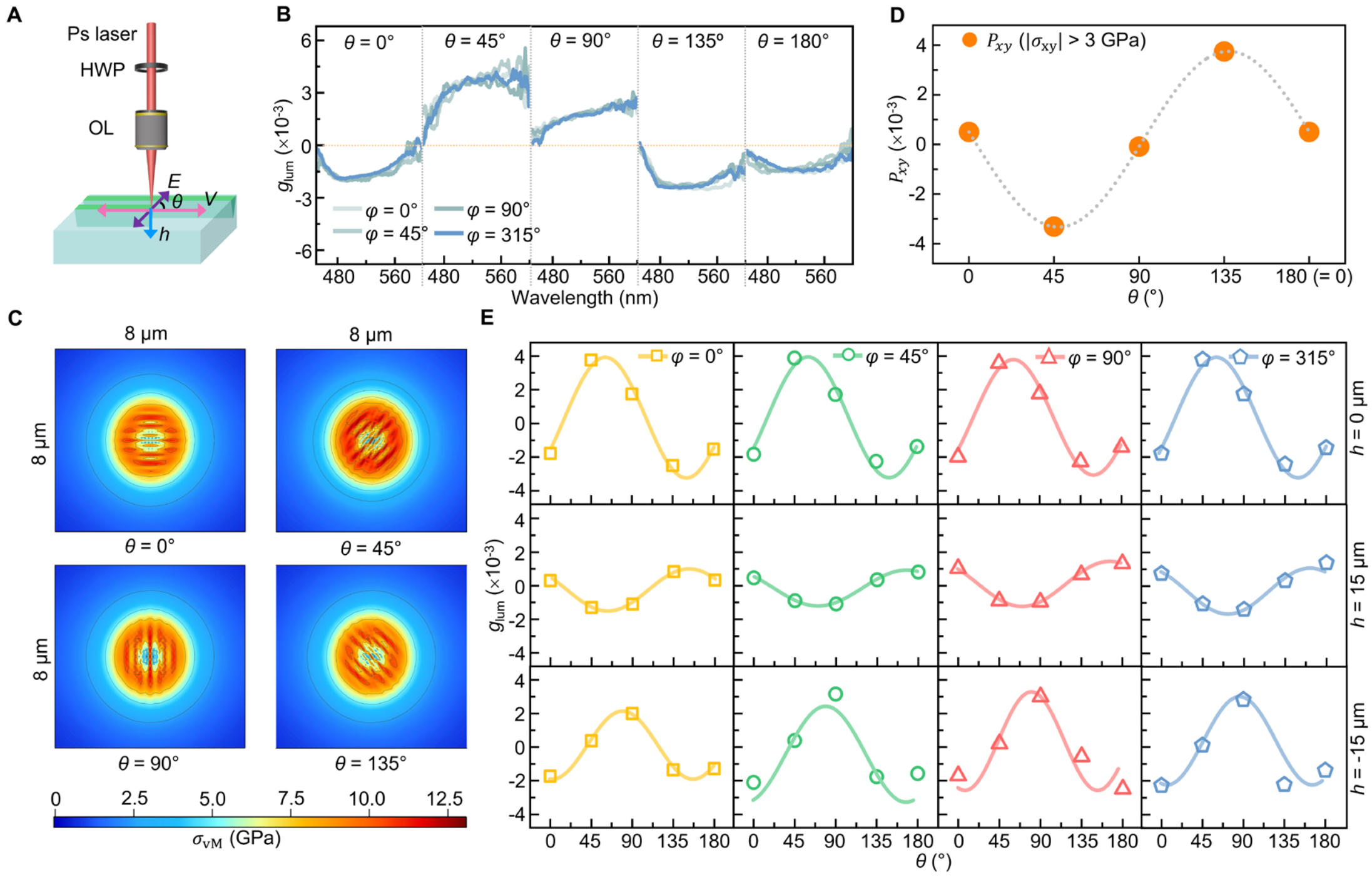

**Figure 4.** Deterministic programming of circularly polarized emission through polarization-dependent laser writing. (A) An illustration of the synthesis parameters: the linear polarization direction of the laser was controlled by a $\lambda/2$ waveplate, where $\theta$ is the angle between the polarization direction ($E$) and the laser-writing axis ($V$); h is the focal depth. (B) Dissymmetry factor ($g_{\mathrm{lum}}$) of the emitted light as a function of $\theta$ and $\varphi$ at h = 0 μm. (C) Simulated distributions of the thermally induced von Mises stress for representative polarization angles $\theta$ = 0°, 45°, 90° and 135° at the laser focus. (D) Finite-element simulation of the normalized signed shear parameter $P_{xy}$ ($|\sigma_{\mathrm{xy}}|$ > 3 GPa) as a function of $\theta$, evaluated over the entire computational domain (20 μm × 20 μm × 37 μm). The gray dotted curve is a sinusoidal fit to the calculated $P_{xy}$ values. (E) $g_{\mathrm{lum}}$ as a function of $\theta$ for different focal depths $h$ and sample azimuthal angles $\varphi$.

We investigated deterministic ways to program the handedness and magnitude of the intrinsic CPL of the perovskite nanocrystals by varying the parameters of laser irradiation, as illustrated in Figure 4A. The linear polarization angle of the incident laser was systematically varied at the laser power density of 63.4 TW/cm$^2$. As illustrated in Figure 4A, $\theta$ is the angle between the laser polarization direction and the laser-writing axis. $h$ is the focal depth of the laser along the z-axis.

For perovskite nanocrystals fabricated at the glass surface ($h$ = 0), the dissymmetry factor of the intrinsic CPL was evaluated at different $\theta$ (0°, 45°, 90°, 135°, and 180°). As shown in Figure 4B, the CPL handedness and the magnitude of the dissymmetry factor are highly dependent on $\theta$. At each $\theta$, the resulting CPL retained the same handedness across all sample azimuthal angles (0°, 45°, 90°, 315°). This polarization dependence suggests that varying $\theta$ modifies the mechanical environment associated with the formation of the nanocrystals.

To clarify the mechanical origin of this angular dependence, we examined the spatial distribution of the laser-deposited energy. Laser irradiation of a silicon-coated titanium film produced a stripe pattern whose orientation varied systematically with the incident polarization (Figure S19). This experimentally determined polarization dependence was incorporated into the finite-element model to define the orientation of the stripe-modulated heat source. The simulated von Mises stress distributions in Figure 4C show that rotation of the stripe-modulated heat source reorganizes the spatial pattern of the highly stressed region. Because von Mises stress is non-signed, the signed *xy*-shear imbalance was quantified separately using the local shear-stress component $\sigma_{xy}$. The stripe-like energy deposition produces a strongly non-uniform shear-stress field consisting of spatially modulated positive and negative regions. The positive and negative shear-stress components differ in their spatial distributions and magnitudes across the simulated region. As $\theta$ is varied, the stripe-like excitation profile and the associated $\sigma_{xy}$ distribution rotate accordingly. This rotation reorganizes the positions, spatial extents, and relative magnitudes of the positive and negative shear-stress regions.

To quantify the signed shear imbalance within the highly stressed regions, we introduced a representative high-stress threshold $\sigma_0$, applied to the local *xy*-shear-stress component $\sigma_{\mathrm{xy}}$, and considered only regions satisfying $|\sigma_{\mathrm{xy}}| > \sigma_0$. The signed *xy*-shear-stress contributions within these high-stress regions were then spatially integrated and normalized to obtain $P_{xy}$, the dimensionless signed shear parameter, at each polarization angle. As shown in Figure 4D, $P_{xy}$, evaluated over the regions satisfying $|\sigma_{\mathrm{xy}}| > 3$ GPa, exhibits a periodic dependence on $\theta$, including alternating positive and negative values. Notably, this angular dependence is similar to the experimentally measured variation in the CPL dissymmetry factor (Figure 4E), including the reversal of the CPL handedness. However, the absolute sign of $P_{xy}$ depends on the adopted coordinate convention and is therefore not directly assigned to a specific CPL handedness. The threshold dependence of $P_{xy}$, including its opposite signs at $\theta = 45°$ and 135°, is further detailed in Figure S20 and Note S3.

The simulations reveal that rotation of the stripe-modulated energy-deposition pattern reorganizes the stress field within the surrounding glass and produces polarization-dependent changes in the signed *xy*-shear imbalance within the strongly sheared regions. In particular, the opposite signs obtained for θ = 45° and 135° indicate opposite signs of the integrated *xy*-shear imbalance under these two writing conditions. Although the continuum model does not explicitly resolve the atomic-scale torsional lattice distortion of individual nanocrystals, the calculated shear imbalance provide a plausible mechanical environment within the surrounding glass that may favor different torsional configurations during confined nanocrystal growth. The calculated angular dependence is therefore consistent with the experimentally observed modulation of the CPL response.

Furthermore, we investigated the influence of focal depth $h$ on the polarization-dependent CPL response by positioning the laser focus above the glass surface (-15 μm), at the surface (0 μm), and within the glass (15 μm) (Figure 4E). At all three focal depths, $g_{\mathrm{lum}}$ retains a periodic dependence on the polarization angle, indicating that polarization-dependent CPL persists throughout the investigated depth range. However, both the amplitude and phase of the angular modulation vary with focal depth. The responses obtained at $h = -15$ μm and $h = 0$ μm exhibit a similar overall handedness sequence, whereas the response at $h = 15$ μm is phase-shifted and displays the opposite CPL sign over corresponding polarization-angle ranges. The observed focal-depth dependence may be associated with changes in the laser focal profile arising from optical astigmatism. Such variations in focal ellipticity are expected to modify the stripe-like energy-deposition pattern and the associated stress distribution, providing a possible origin of the observed phase shift in the polarization-dependent CPL response.

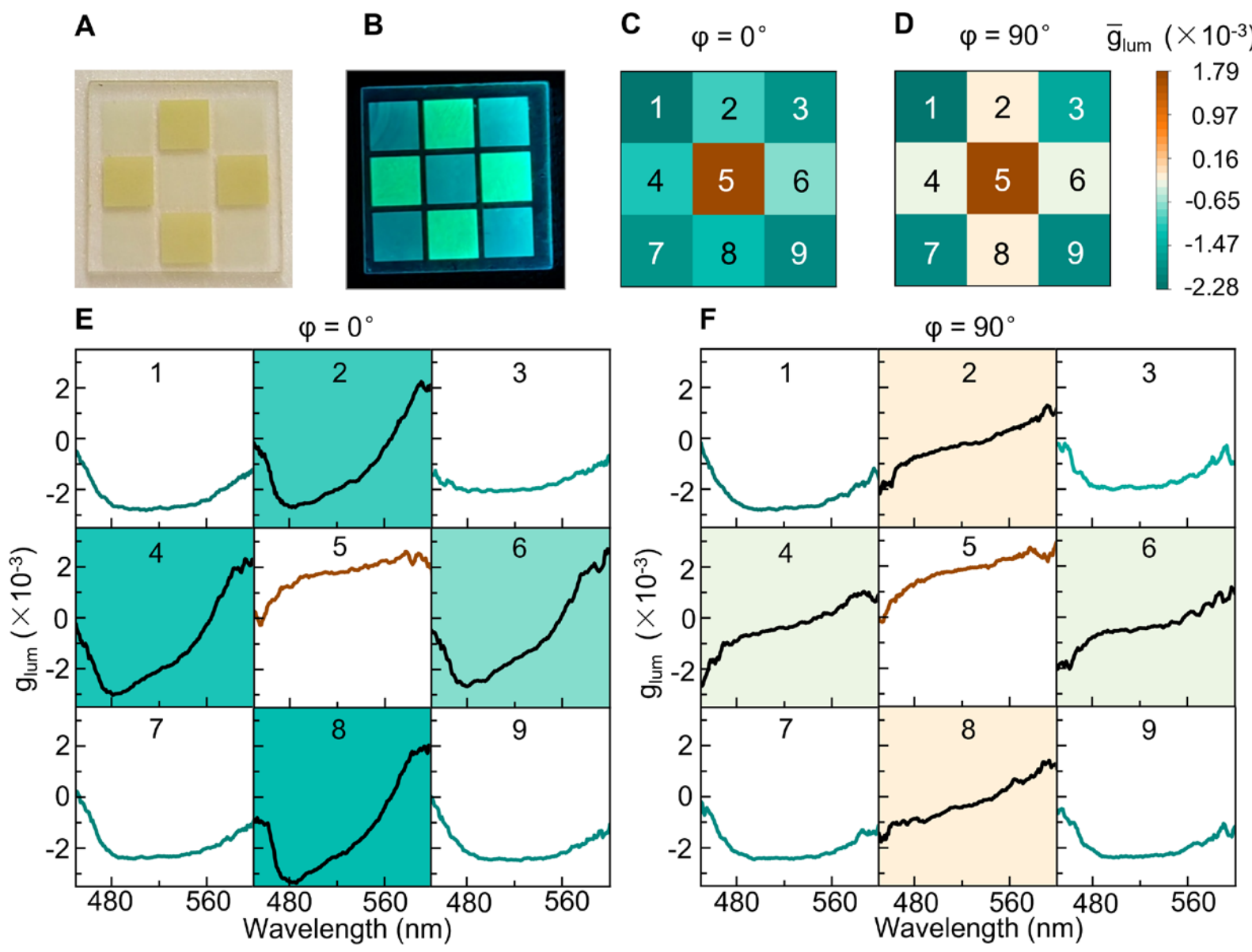

**Figure 5.** Multichannel optical information encoding based on programmable chiroptical responses. Images of the sample under (A) daylight and (B) 365 nm excitation. Color maps of the wavelength-averaged luminescence dissymmetry factor $\overline{g}_{\mathrm{lum}}$, obtained by averaging the measured $g_{lum}$ spectra over 450-600 nm, at sample azimuthal angles of (C) $\varphi = 0°$and (D) $\varphi = 90°$. Corresponding measured $g_{lum}$ versus wavelength from nine regions of the sample at (E) $\varphi = 0°$and (F) $\varphi = 90°$. The glass sample shown in (A) and (B) was a square piece with dimensions of approximately 1 cm × 1 cm.

We designed a proof-of-concept multichannel optical encoding pattern based on these programmable features. A 3 × 3 array was fabricated within the glass matrix by independently varying the laser power density and the linear polarization angle $\theta$. Following thermal annealing at 350 °C for 10 h, the array exhibits laser power density-dependent optical responses, enabling information encoding through three optical readout channels.

The first two encoding layers rely on macroscopic visual identifiers. Regions processed with a high laser power density (126.9 TW/cm$^2$; regions 2, 4, 6, 8) appear as yellow squares under daylight (Figure 5A) and exhibit bright green fluorescence under 365 nm UV excitation (Figure 5B). Conversely, regions processed at 63.4 TW/cm$^2$ (regions 1, 3, 5, 7, 9) remain visually transparent under daylight (Figure 5A) and emit cyan photoluminescence under 365 nm UV excitation (Figure 5B). Details of the fabrication are included in Table S5.

The third encoding layer utilizes chiroptical signatures and is read out by azimuthal-angle-resolved CPL spectroscopy. To resolve this layer, we extracted the dissymmetry factors across the 450-600 nm emission range at two azimuthal angles $\varphi$ (0° and 90°) (Figures 5C-5F). For the 63.4 TW/cm$^2$ regions, the CPL spectra retain the same handedness at both azimuthal angles, consistent with the intrinsic CPL established above. In contrast, the 126.9 TW/cm$^2$ regions exhibit strongly rotation-dependent CPL signals associated with linear optical anisotropy. As shown in Figures 5E and 5F, the measured $g_{\mathrm{lum}}$ spectra of these regions change substantially between $\varphi = 0°$ and 90°. The distinct azimuthal dependences of these two responses therefore provide an additional chiroptical dimension for encoding and resolving spatially patterned optical information.

## Conclusion

In summary, we demonstrate deterministic programming of intrinsic CPL from laser-written nanocrystals, with its origin associated with lattice distortion and band-edge electronic structure. The observed lattice distortion can be described by a torsional-distortion model that breaks inversion symmetry, while DFT calculations show that the distortion lifts the spin degeneracy of the band-edge electronic states. By varying the writing conditions, we deterministically control the handedness and magnitude of the intrinsic CPL. We further use this programmability to encode multiple optical channels within a monolithic glass.

The ability to write within a transparent bulk material could enable three-dimensional chiroptical architectures in which emission handedness is encoded throughout the volume. Extending reversible laser writing to the control of lattice distortion could enable dynamic switching of CPL handedness after fabrication. At the single-emitter limit, such control may provide a route towards structurally programmable chiral spin-photon interfaces. More broadly, direct writing could evolve from defining where a light source is formed to controlling the structural symmetry and electronic states from which light is emitted.

## Experimental Section

### Material fabrication

The melting-quenching method was employed to prepare the precursor glass. The composition of the original reagent-grade powder was 21$SiO_2$ (99.99%)-53.2$B_2O_3$ (98%)-6ZnO (99.9%)-3.8$SrCO_3$ (99.95%)-7$Cs_2CO_3$ (99.9%)-3PbO (99.9%)-6NaBr (99.9%). Approximately 100 g of raw powder was weighed and uniformly mixed for at least 20 min. Then the powder was transferred to an alumina crucible for melting at 1000 °C for 60 min. After thorough melting, the glass melt was poured onto a room-temperature graphite plate. Following rapid quenching by compression between stainless steel plates, the glass was annealed at 350 °C for 10 h in a muffle furnace. Finally, the precursor glass was fabricated into specimens of defined dimensions (10 mm × 10 mm × 1.5 mm) prior to laser irradiation and heat treatment.

### Laser direct writing

The ultrafast laser direct writing setup was equipped with a diode-pumped fiber amplifier laser system (Tangerine, Amplitude), which employed ytterbium-doped photonic crystal fibers as an amplifying medium. The laser wavelength was 1030 nm, the pulse duration was set to 5 ps, and the repetition frequency was set to 10 kHz. The laser beam was focused onto the glass sample surface through a 0.3-NA objective lens. The sample was positioned on a three-dimensional computer-controlled translation stage with the fixed scanning speed of 300 μm/s and the parallel pitch of 30 µm. A half-wave plate was placed in the optical path to alter the laser linear polarization direction. The glass samples were finely polished prior to laser writing. The laser power was controlled by an attenuator, and the number of pulses for the single-spot irradiation experiment was jointly regulated by shutter opening/closing time and laser frequency setting. The laser power ranged from 100 to 1000 mW. Prior to each sample preparation, the power transmittance through 0.3-NA objective was measured and consistently found to be 36%, ensuring accurate determination of the laser power delivered to the sample. The patterned regions were written using a bidirectional serpentine raster, in which adjacent lines were traversed in opposite directions. Throughout this work, $V$ denotes the unoriented laser-writing axis rather than the signed translation velocity. After laser irradiation, annealing was performed at 350 °C for 10 h to facilitate the growth of $CsPbBr_3$ NCs in the sample.

**Circularly polarized photoluminescence characterization**

Circularly polarized photoluminescence measurements were performed using a JASCO CPL-300 spectrometer with excitation at 365 nm. The emission spectra were recorded from 450 to 600 nm with a wavelength interval of 1 nm. The emitted light was analyzed using a photoelastic modulator (PEM) operating at 50 kHz and a linear polarizer fixed at 45° relative to the PEM optical axis, and detected using a photomultiplier tube (PMT) (Figure S4). For azimuth-dependent measurements, the sample was rotated about the optical axis to the specified azimuthal angle φ while the excitation and detection configuration remained fixed. For all measurements, the luminescence DC signal was adjusted to 0.5 V using the high-voltage control loop.

**TEM Characterization**

HRTEM images were acquired using a 200 kV transmission electron microscope (Talos F200X G2, Thermo Fisher Scientific) at an electron dose rate of 326 e/(Å$^2$·s), with a total electron dose of approximately 19,560 e/Å$^2$ per image.

**Finite-element simulations**

Three-dimensional transient thermo-mechanical finite-element simulations were performed using COMSOL Multiphysics. Transient heat transfer was coupled to quasi-static solid mechanics through thermal expansion, with the glass treated as a homogeneous isotropic continuum. For the power-dependent residual-stress simulations, laser energy deposition was represented by a moving volumetric Gaussian heat source with exponential depth decay, and the power-dependent absorption coefficients were derived from the experimentally measured absorbance. Structural relaxation of the glass was phenomenologically described using a six-branch Generalized Maxwell model combined with the Williams-Landel-Ferry time-temperature superposition principle. For the polarization-dependent simulations, a stripe-modulated elliptical Gaussian volumetric heat source was used to evaluate the resulting spatial distribution of the laser-induced $xy$-shear stress. Detailed model parameters, boundary conditions, meshing, solver settings, and post-processing procedures are provided in Supplementary Notes S1 and S3.

**DFT calculations**

First-principles calculations were performed in the framework of density functional theory (DFT) with plane-wave basis sets, periodic boundary conditions, and Norm-Conserving pseudopotentials as implemented in the GPU-accelerated PWmat package.[26-30] Exchange-correlation effects were treated within the generalized gradient approximation (GGA) using the Perdew-Burke-Ernzerhof (PBE) functional.[31] The maximum cutoff energy for the plane wave basis sets was 50 Ry. To account for long-range van der Waals interactions, the DFT-D3 correction was applied,[32] and spin-orbit coupling (SOC) was included in all simulations.[33] The Brillouin zone was sampled using a Monkhorst-Pack grid[34] with dimensions of $4 \times 4 \times 4$. The reference cubic structure was first optimized until the force on every atom was below 0.02 eV/Å before performing band structure calculations. The torsionally distorted models were subsequently constructed from the optimized structure by imposing the prescribed atomic rotations. For these torsional models, the lattice constants were relaxed while the imposed atomic positions were retained before band-structure calculations.

## Acknowledgments

This work was supported by the National Natural Science Foundation of China (12674398, 12504024), the Science and Technology Commission of Shanghai Municipality (25DZ3001902), and SOE-DSPF of Westlake University. We acknowledge Yuan Cheng, Xue Lou, Zhong Chen, Lingyu Xiao, Xiaohe Miao, Wentian Song, Pei Sheng, and Jingli Zhao for technical assistance; and the Instrumentation and Service Center for Molecular Sciences (ISCMS), Instrumentation and Service Center for Physical Sciences (ISCPS), and Westlake Center for Micro/Nano Fabrication (WCMNF) at Westlake University for facility support.

## Author contributions

S.P. supervised the project. Y.J. and S.P. conceived and designed the experiments. Y.J. fabricated the samples and performed the optical measurements, with assistance from X.S., W.L., J.D., Z.Z., P.Y. and B.S. Z.L. performed the transmission electron microscopy characterization. Y.C. and C.C. performed the density functional theory calculations, and Y.H. performed the COMSOL simulations. Y.J., Y.C., Y.H., S.P. and Y.W. analyzed the data. Y.J. and S.P. wrote the manuscript, with revisions from M.Q. and Y.W. All authors reviewed and approved the final manuscript.

# Contents

**Note S1.** Phenomenological simulations of power-dependent residual stress.

The laser power density reported throughout this work, I, was defined as $I = \frac{P_0}{\pi\omega_0{}^2}$, where $P_0$ is the peak power of a single pulse and $\omega_0 = 1.202\ \mu m$ is the Gaussian beam radius. In the COMSOL heat-source model, the spatial intensity distribution was described by a Gaussian profile, for which the on-axis peak power density is $I_0 = \frac{2P_0}{\pi\omega_0{}^2} = 2I$. Therefore, the experimentally defined laser power density of 63.4 TW/cm$^2$ corresponds to an on-axis Gaussian peak power density of 126.8 TW/cm$^2$ in the simulation. For the 63.4 TW/cm$^2$ condition, the measured power after the 10× objective was 144 mW at a repetition rate of 10 kHz and a pulse duration of 5 ps, corresponding to a pulse energy of 14.4 μJ and a peak pulse power of 2.88 MW.

To elucidate the non-monotonic dependence of residual stress on laser power density, a three-dimensional thermo-mechanical coupled finite element model was implemented in COMSOL Multiphysics to examine the relative evolution of the laser-induced residual-stress response with laser power density. The transient temperature field was calculated utilizing a moving volumetric heat source characterized by a spatial Gaussian profile and an exponential depth decay. The power-dependent absorption coefficient was derived from experimental absorbance measurements (Figure S10A) to quantify the nonlinear energy deposition. The resulting thermal expansion was coupled to a quasi-static solid mechanics module.

To phenomenologically describe temperature-accelerated structural relaxation in the laser-modified glass, a six-branch Generalized Maxwell viscoelastic model was employed in conjunction with the Williams-Landel-Ferry (WLF)[1] time-temperature superposition principle (constants 17.44 and 51.6 K). Based on the experimentally determined glass transition temperature (728.15 K)[2] and previous reports showing that laser irradiation can substantially modify the local mechanical properties of glass,[3,4] an effective long-term shear modulus of $2.0 \times 10^7$ Pa was adopted for the laser-modified region as a phenomenological model parameter. The model was then used to describe the transient structural relaxation and subsequent freezing of thermal strain during rapid cooling.

The simulations yield a distinct intermediate power density transition regime (31.7–63.4 TW/cm$^2$) associated with a relatively high residual-stress response at the sampled location after the modeled cooling interval (Figure S10B). At intermediate power densities, the rapid cooling process outpaces viscous relaxation, freezing a relatively large anisotropic thermal strain into the lattice. This relatively high simulated residual stress aligns with the experimental realization of the core-shell nanocrystal morphology and the emergence of intrinsic chiral emission. Conversely, at higher power densities, the localized peak temperature substantially exceeds the glass transition temperature. In this elevated temperature regime, accelerated viscoelastic relaxation allows the accumulated thermal strain to dissipate prior to structural freezing. This simulated stress-relieved state corroborates experimental observations at high power densities: absence of core-shell boundaries in TEM (Figure S11) and the absence of intrinsic circularly polarized photoluminescence.

Material parameters used for the simulation are summarized in Table S2. Generalized Maxwell model parameters are summarized in Table S3. The initial temperature of the model was set to 293.15 K. A thermal-insulation condition was applied to the model boundaries, with a periodic thermal condition imposed between boundaries 2 and 7. In the solid-mechanics module, boundaries 2 and 7 were fixed in all three displacement components to suppress rigid-body motion, while the remaining boundaries were treated as mechanically free. The mesh was constructed using free-triangular and swept prismatic elements, resulting in 29,600 prism elements with a minimum element quality of 0.4965 and an average element quality of 0.7921. Local refinement was applied near the laser-interaction region with a maximum element size of $r_0/8$, and 20 swept layers were used along the depth direction. The transient simulation was performed from 0 to 0.4 ms. To isolate the thermo-mechanical response to a single laser pulse, the recurrence interval of the explicit heat-source events was set to 100/f=10 ms, which is much longer than the simulated time window. Therefore, only one 5 ps heat pulse was applied during each simulation, followed by post-pulse cooling and stress relaxation. A BDF time-stepping scheme with a maximum order of 2, an initial time step of 1 fs, and a relative tolerance of 0.001 was used, and the linear systems were solved using the PARDISO direct solver. For Figure S10B, the residual von Mises stress was extracted at (x, y, z) = (0, 0, -1 μm) at the end of the simulated interval (t = 0.4 ms) for each laser-power condition.

**Note S2.** DFT simulations of torsionally distorted $CsPbBr_3$.

The bond-length distortion index ($D$), bond-angle variance ($\sigma^2$), and the disparity in the Pb-Br-Pb bond angle ($\beta$), denoted as $\Delta\beta$, were calculated using the following equations and summarized in Table S4:

$$D = \frac{1}{6}\sum_{i=1}^{6}\frac{|d_i - d_0|}{d_0} \quad \text{(S1)}$$

$$\sigma^2 = \frac{1}{11}\sum_{i=1}^{12}(\theta_i - 90°)^2 \quad \text{(S2)}$$

$$\Delta\beta = \beta_{\max} - \beta_{\min} \quad \text{(S3)}$$

where $d_i$ is the corresponding Pb-Br bond distance, $d_0$ represents the average Pb-Br bond length, and $\theta_i$ denotes the Br-Pb-Br bond angle within the $PbBr_6$ octahedron. To quantitatively evaluate the influence of twisting distortion on the local octahedral geometry of cubic $CsPbBr_3$, we analyzed the structural deformation under the untwisted condition, at the twisting angles of 1.0° and 2.8°. For each distorted model, the lattice constants were relaxed while the atomic positions were kept fixed. The structural distortion was evaluated using three key structural parameters: the bond-length distortion index ($D$), the bond-angle variance ($\sigma^2$), and the in-plane $\beta$ angle disparity ($\Delta\beta$). The untwisted cubic structure is essentially undistorted: $D$ = 3.68 × $10^{-7}$ and $\sigma^2$ = 8.71 × $10^{-5}$ degree$^2$, with $\beta$ approximately 180.0° and $\Delta\beta$ approximately 0.0°. As the maximum imposed twisting angle increases from 1.0° to 2.8°, $D$ increases from 3.78965 × $10^{-5}$ to 2.44218 × $10^{-4}$, while $\sigma^2$ increases from 0.363870 to 2.849912. The corresponding unrounded $\Delta\beta$ values are 0.4149° and 1.1610° for the 1.0° and 2.8° twisted structures, respectively. The 2.8° model therefore has the largest distortion, but bond-length distortion remains small and angular deformation dominates.

**Note S3.** Polarization-dependent thermo-mechanical simulations and signed *xy*-shear analysis.

To examine the polarization-dependent thermo-mechanical response, a three-dimensional transient finite-element model was established using COMSOL. The threshold-dependent shear analysis presented below was performed at t = 10 ps over the entire computational domain $\Omega$ (20 μm × 20 μm × 37 μm). Transient heat transfer was coupled to quasi-static solid mechanics through thermal expansion, with the glass treated as a homogeneous isotropic continuum. A stripe-modulated elliptical Gaussian volumetric heat source was used to represent the laser-induced energy deposition. An angle-dependent normalization factor was applied to the stripe-modulated heat source to maintain a consistent spatially integrated heat-source amplitude upon rotation of the stripe orientation. Simulations were performed for polarization-writing-axis angles of 0°, 45°, 90° and 135°, while all other model parameters were kept unchanged.

The laser wavelength, pulse duration, repetition rate, and scanning speed were 1030 nm, 5 ps, 10 kHz, and 300 μm/s, respectively. The laser power after the objective was 0.144 W, corresponding to the experimentally defined laser power density of 63.4 TW/cm$^2$ and an on-axis Gaussian peak power density of 126.8 TW/cm$^2$ in the simulation. The stripe period and modulation depth were 0.4 μm and 0.3, respectively. The material parameters used in the model were $\rho$=3800 kg/m$^3$, $k$=1.1 W/(m·K), $C_p$ = 650 J/(kg·K), $\alpha_T$=6.0 × 10$^{-6}$ 1/K, $K$=4.286 × 10$^{10}$ Pa and $G$ = 2.952 × 10$^{10}$ Pa. Representative thermal and elastic properties were used as effective model parameters.

The initial temperature was set to 293.15 K. Selected external boundaries were maintained at 293.15 K, boundaries 2 and 15 were coupled by a periodic thermal condition, and the other external boundaries were thermally insulated. In the solid-mechanics model, $u_y = 0$ was imposed on boundaries 2 and 15, $u_x = 0$ on boundaries 1, 5, 9, 11, and 16-18, and $u_z = 0$ on boundaries 3, 7, and 13, while the remaining displacement components were left free. The mesh was locally refined in the laser-interaction region to a target element size of 0.0667 μm, resulting in 72,450 elements with a minimum element quality of 0.5326. Transient simulations were performed from 0 to 1 ms using a BDF time-stepping scheme with a relative tolerance of 1 × 10$^{-4}$. The results were sampled at 5 ps intervals during the first 1 ns and at approximately 5 μs intervals thereafter.

The normalized signed *xy*-shear parameter within the threshold-selected region is denoted by $P_{xy}(\sigma_0)$, which is dimensionless and ranges from −1 to 1.

$$P_{xy}(\sigma_0)=\frac{\int_{\Omega} H(|\sigma_{\mathrm{xy}}|-\sigma_0)\sigma_{xy}\mathrm{d}V}{\int_{\Omega} H(|\sigma_{\mathrm{xy}}|-\sigma_0)|\sigma_{xy}|\mathrm{d}V} \tag{S4}$$

where the selection function $H$ is given by

$$H(|\sigma_{\mathrm{xy}}| - \sigma_0) = \begin{cases} 1, & |\sigma_{\mathrm{xy}}| > \sigma_0, \\ 0, & |\sigma_{\mathrm{xy}}| \le \sigma_0. \end{cases} \tag{S5}$$

Here, $\sigma_{\mathrm{xy}}$ is the local *xy*-shear-stress component, $\sigma_0$ is the selected stress threshold, and $\Omega$ denotes the entire computational domain. The numerator represents the net signed xy-shear contribution within the region satisfying $|\sigma_{\mathrm{xy}}| > \sigma_0$, whereas the denominator represents the total absolute xy-shear contribution within the same region. Therefore, the sign of $P_{xy}$ indicates the net sign of the *xy*-shear imbalance, while its magnitude reflects the degree of imbalance between the positive and negative shear contributions.

As shown in Figure S20, at low values of $\sigma_0$, regions with relatively small $|\sigma_{\mathrm{xy}}|$ occupy a large fraction of the simulated volume, and their positive and negative shear contributions largely cancel. With increasing $\sigma_0$, the analysis becomes progressively restricted to the strongly sheared regions, revealing a more pronounced angular dependence of $P_{xy}$. When $P_{xy}$ is evaluated over the regions satisfying $|\sigma_{\mathrm{xy}}|$ > 3 GPa, the opposite signs obtained for θ = 45° and 135° indicate small but oppositely signed net *xy*-shear imbalances under these two writing configurations. Accordingly, $\sigma_0$ = 3 GPa was selected as a representative post-processing threshold, because the directional imbalance of the *xy*-shear component becomes clearly resolved at this threshold. The corresponding $P_{xy}$ values used in Figure 4D serve as a comparative metric rather than a critical stress for lattice deformation.

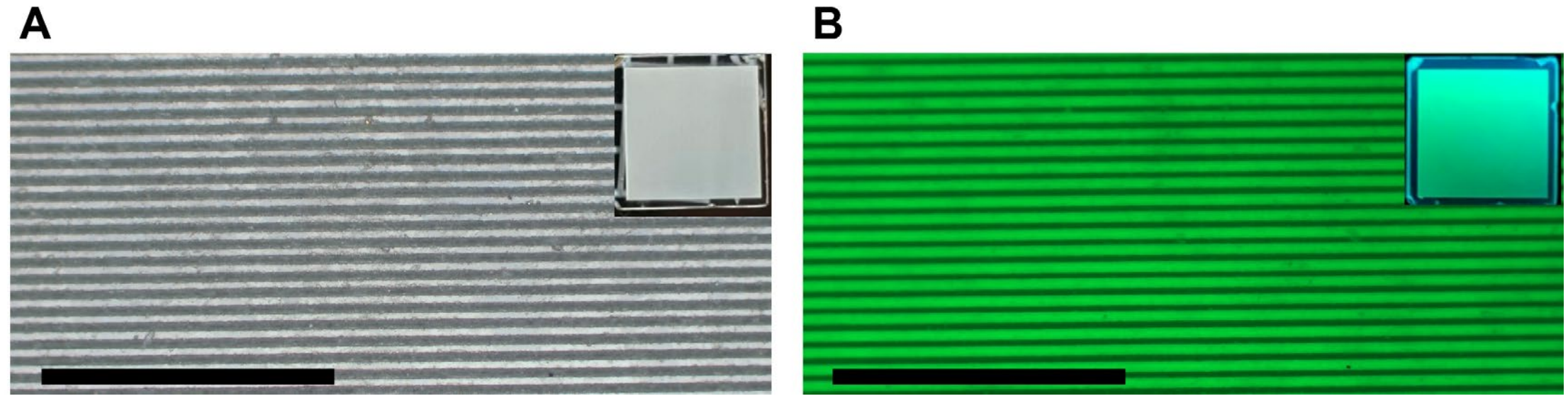

**Figure S1.** Optical characterization of the laser-written sample. The sample was written at a laser power density of 63.4 TW/cm$^2$. Optical micrographs acquired under white-light illumination (A) and 405 nm laser excitation through a 0.3-NA objective (B), respectively. Insets: photographs of the corresponding sample under daylight and 365 nm UV illumination. Scale bars, 500 μm. The glass sample shown in (A) and (B) was a square piece with dimensions of approximately 1 cm × 1 cm.

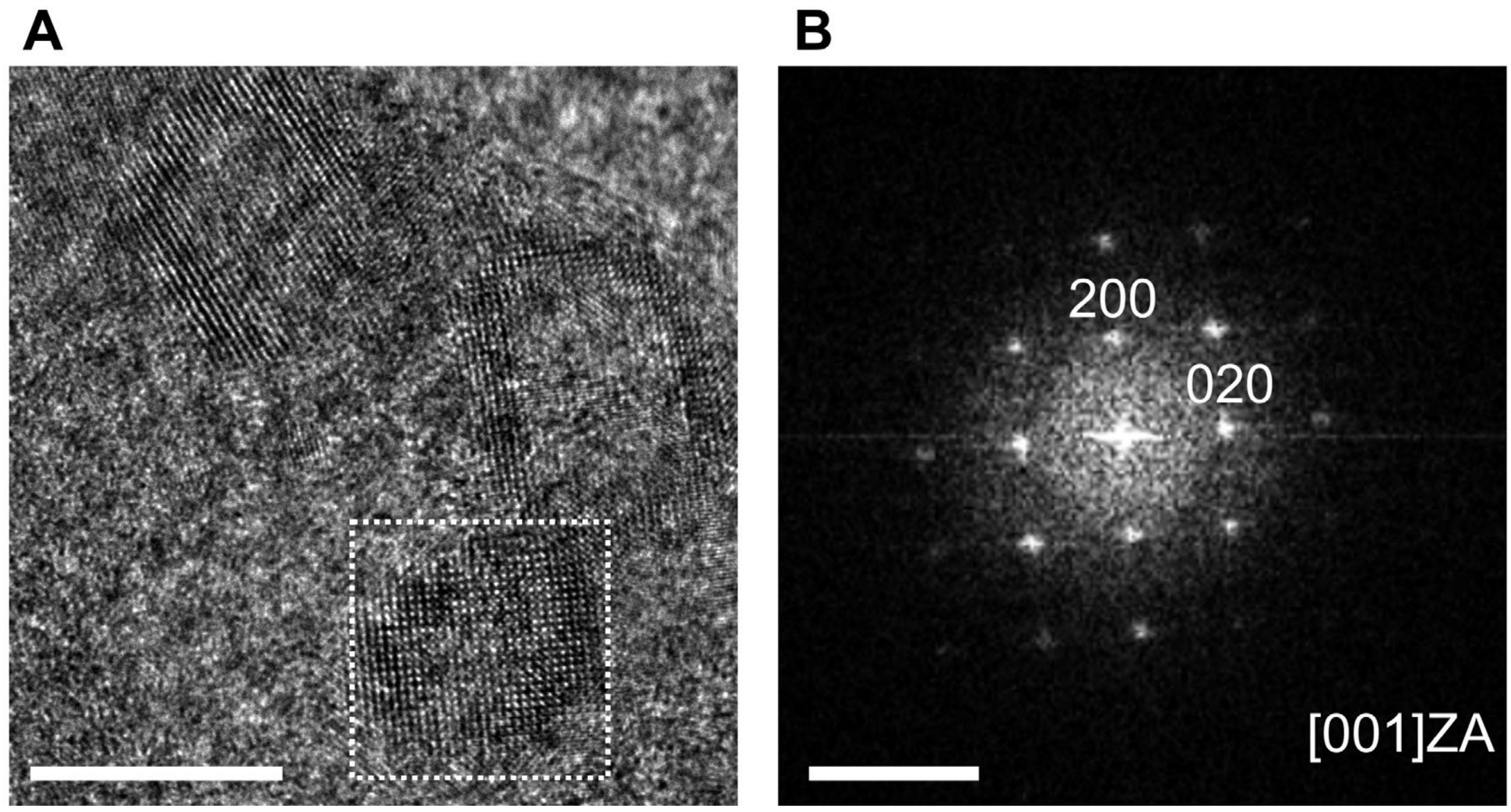

**Figure S2.** (A) HRTEM image of core-shell PNCs embedded in the glass. The region enclosed by the white dashed box is enlarged in Figure 1B. The sample was fabricated at a laser power density of 63.4 $TW/cm^2$. Scale bar, 10 nm. (B) Corresponding FFT pattern of the white dashed boxed PNC viewed along the [001] zone axis. Scale bar, 5 1/nm.

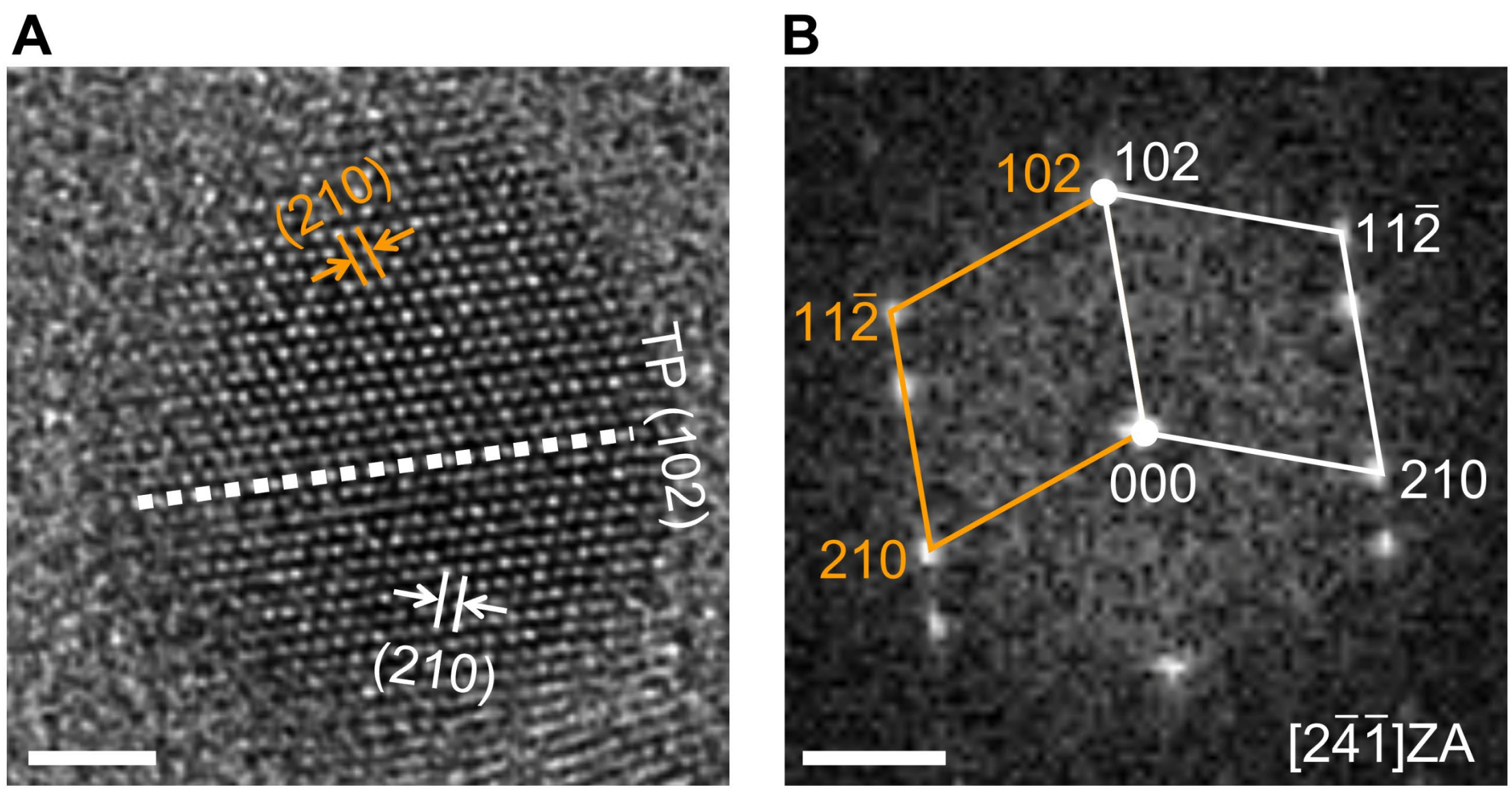

**Figure S3.** HRTEM characterization of a twinned PNC. (A) The twin plane (TP), indicated by the white dashed line, is indexed as (102). Scale bar, 2 nm. (B) The corresponding FFT pattern shows diffraction spots from the two twin-related domains viewed along the $[2\bar{4}\bar{1}]$ zone axis. Scale bar, 2 1/nm.

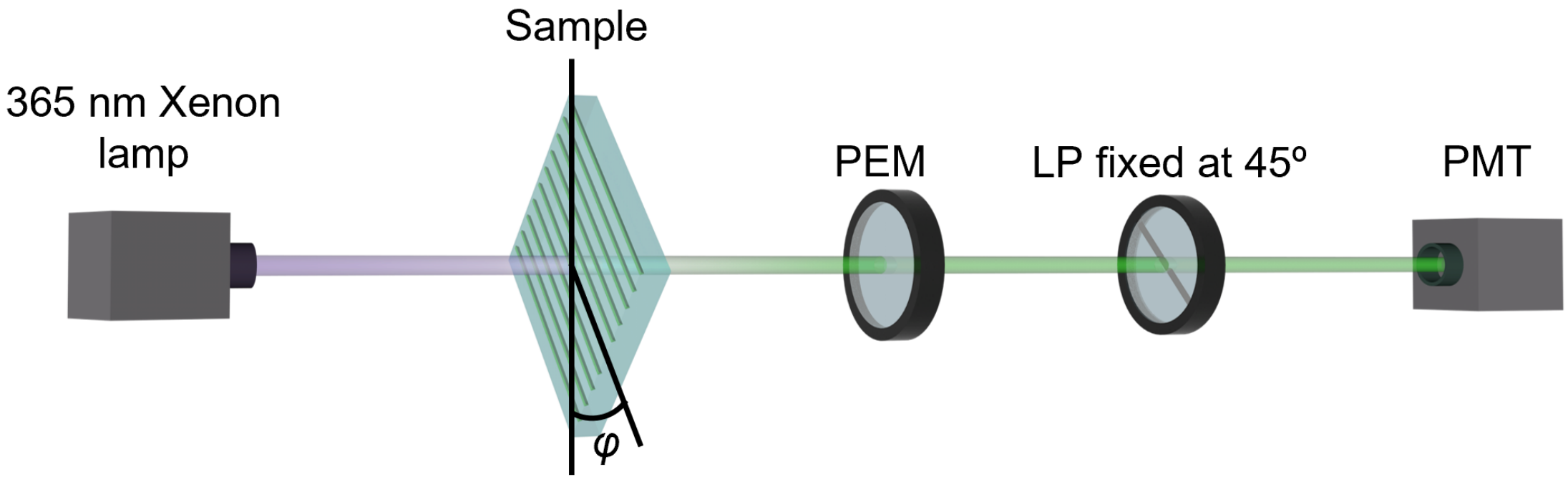

**Figure S4.** Schematic of the azimuth-dependent CPL measurement setup. The sample was excited at 365 nm using a xenon lamp in a JASCO CPL-300 spectrometer. The emitted light was analyzed using a photoelastic modulator (PEM) operating at 50 kHz and a linear polarizer (LP) fixed at 45° relative to the PEM optical axis, and detected using a photomultiplier tube (PMT). The angle φ denotes the sample azimuthal angle during rotation about the optical axis.

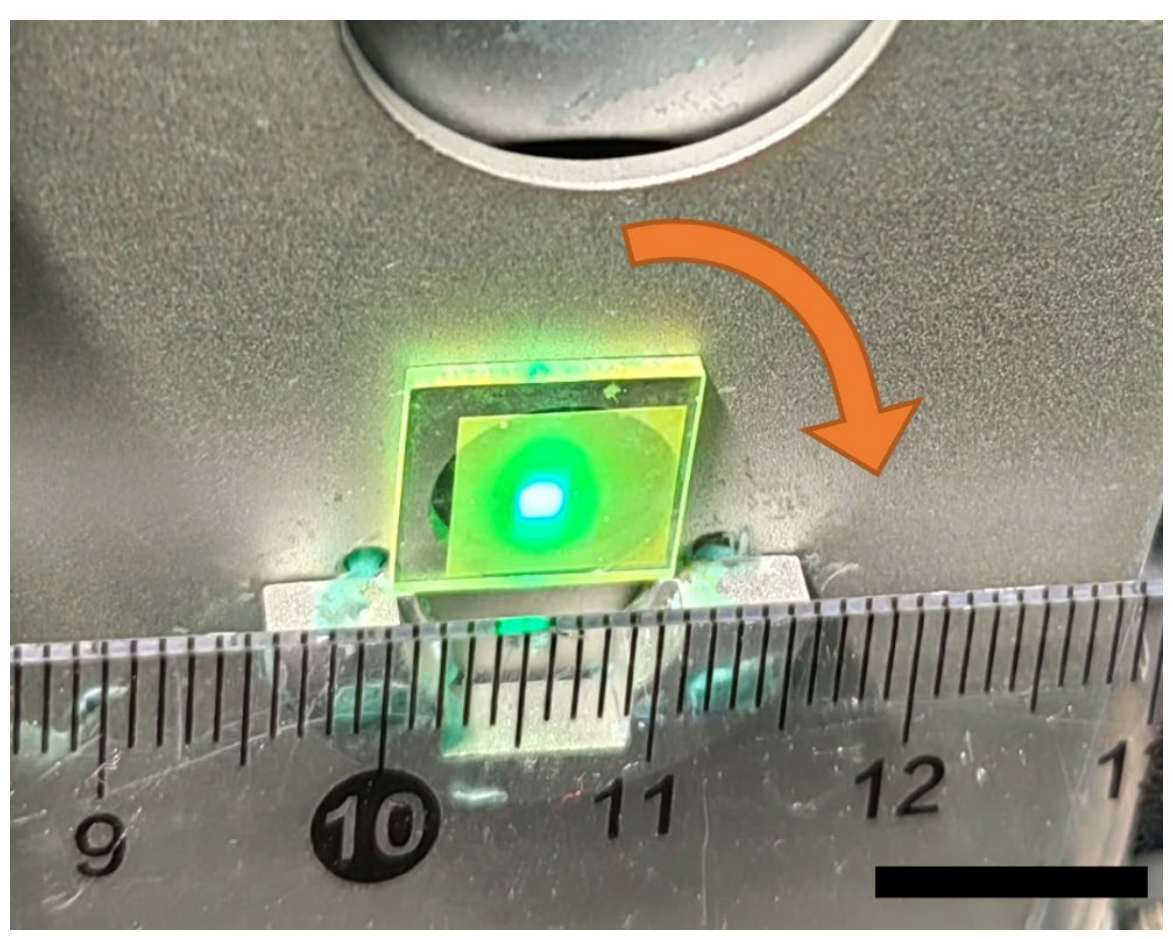

**Figure S5.** CPL measurement configuration and excitation area. The excitation spot size on the sample was approximately 1.7 mm × 1.7 mm. The orange arrow indicates the direction of sample rotation around the optical axis during the azimuth-dependent CPL measurements. Scale bar, 1 cm.

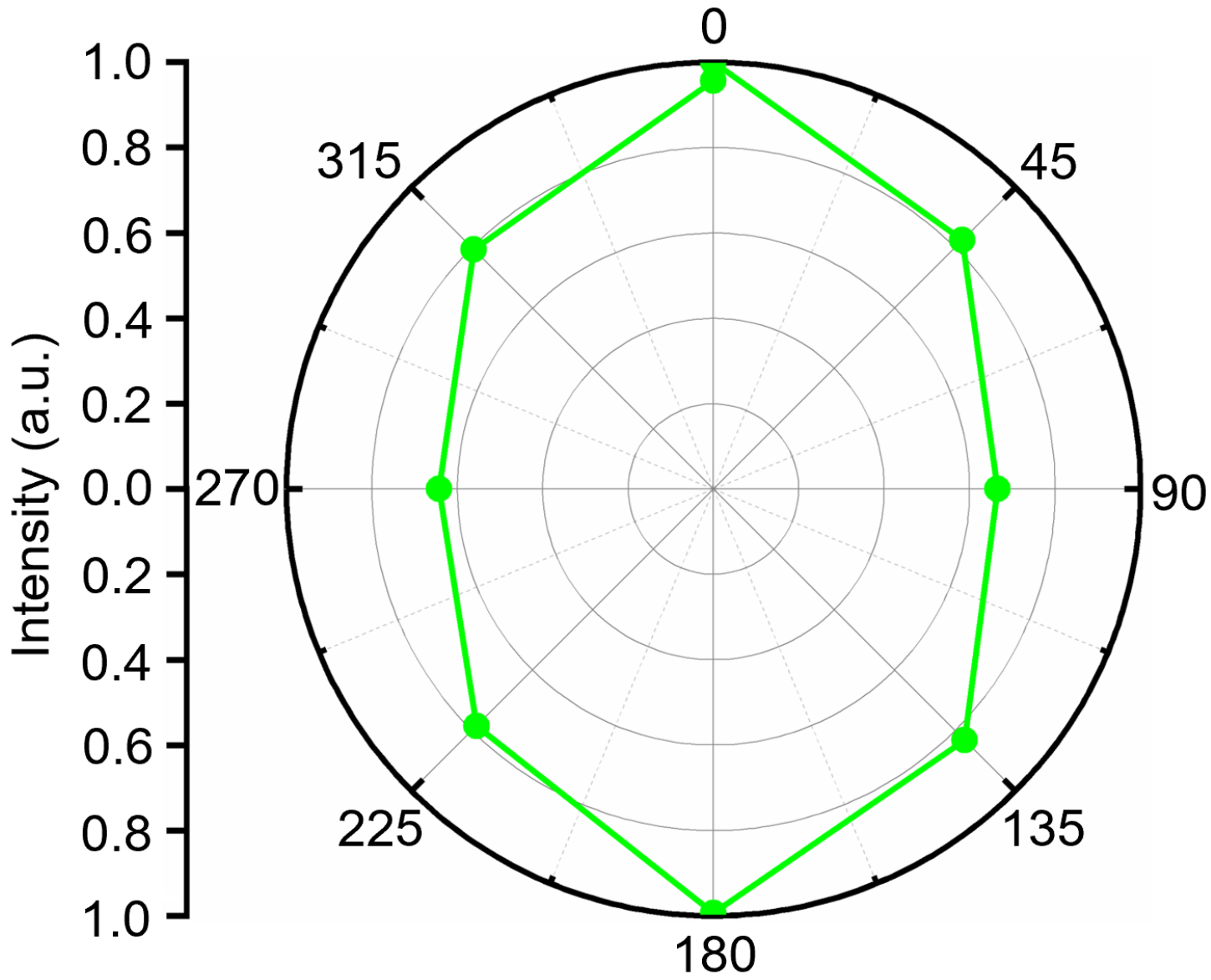

**Figure S6.** Polarization-resolved photoluminescence of the laser-written sample. Polar plot of the normalized PL intensity as a function of the analyzer angle, obtained by rotating the linear polarizer in the emission-detection path while keeping the excitation conditions, excitation polarization, and sample orientation unchanged. The sample was excited using a 450 nm continuous-wave laser with a power of 13 μW through a 50× objective. The sample was fabricated at a laser power density of 63.4 TW/cm$^2$, followed by low-temperature heat treatment at 350 °C for 10 h. The angular dependence indicates a linear-polarization component in the emitted light. The degree of linear polarization was calculated according to

$$DOLP = \frac{I_{max} - I_{min}}{I_{max} + I_{min}} \tag{S6}$$

where $I_{max}$ and $I_{min}$ are the maximum and minimum PL intensities measured during analyzer rotation, respectively. The calculated DOLP is 0.1976, indicating a linear-polarization component in the emitted light. This value agrees well with the DOLP of 0.1949 obtained from the independent Stokes-polarimetry measurement in Table S1.

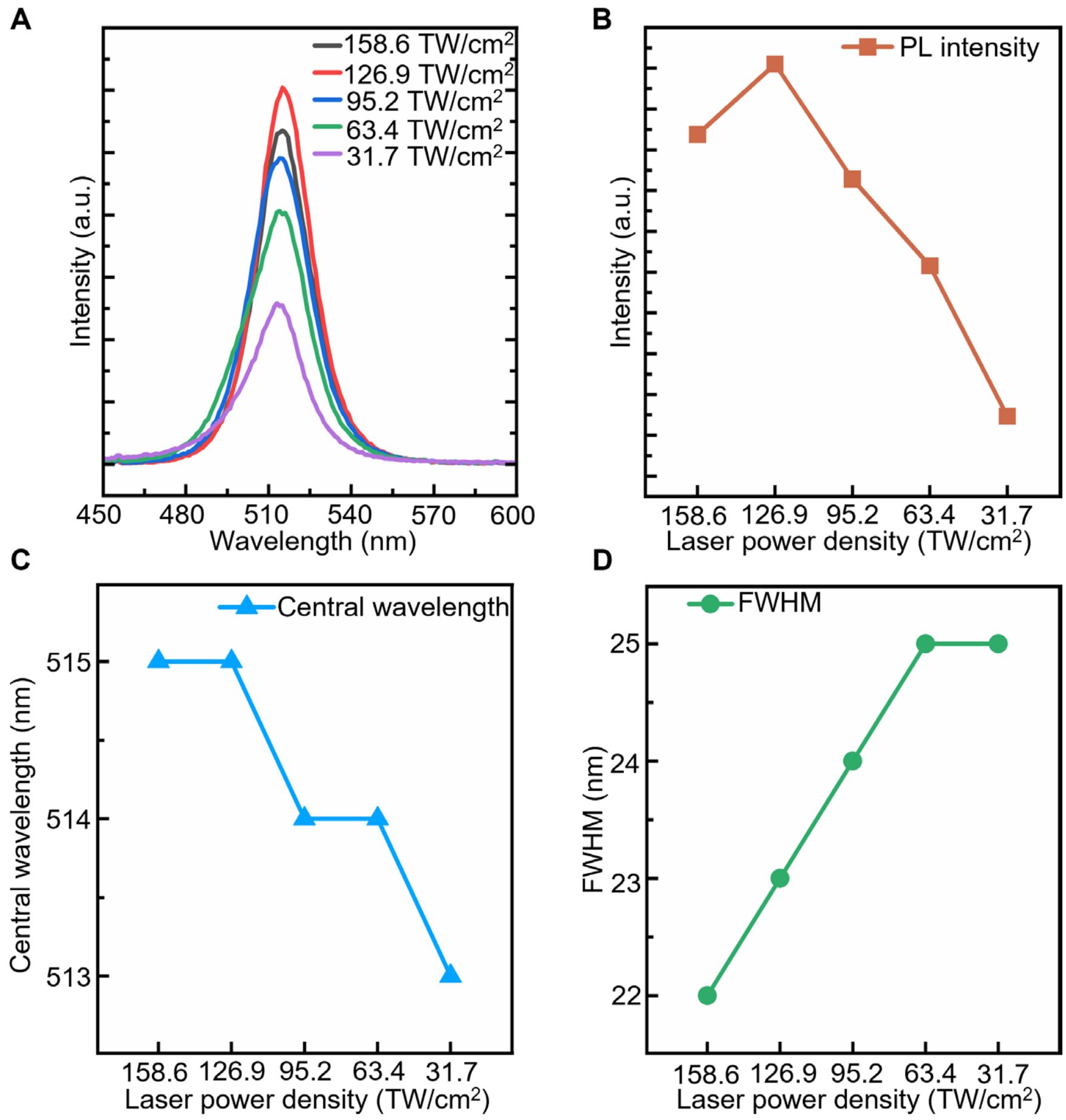

**Figure S7.** Power-dependent photoluminescence of the laser-written sample with heat treatment. The samples were first fabricated by laser writing at power densities of 158.6, 126.9, 95.2, 63.4, and 31.7 TW/cm$^2$, followed by low-temperature heat treatment at 350 °C for 10 h. (A) PL emission spectra of the samples. (B) Corresponding peak PL intensity, (C) emission peak central wavelength, and (D) full width at half maximum (FWHM), respectively. The PL spectra were measured using an Edinburgh Instruments FLS1000 fluorescence spectrometer under 365 nm excitation. The emission spectra were recorded from 450 to 600 nm with a wavelength interval of 1 nm and a dwell time of 0.5 s. The excitation and emission bandwidths were 0.50 and 0.30 nm, respectively, with excitation-reference and emission-response corrections enabled.

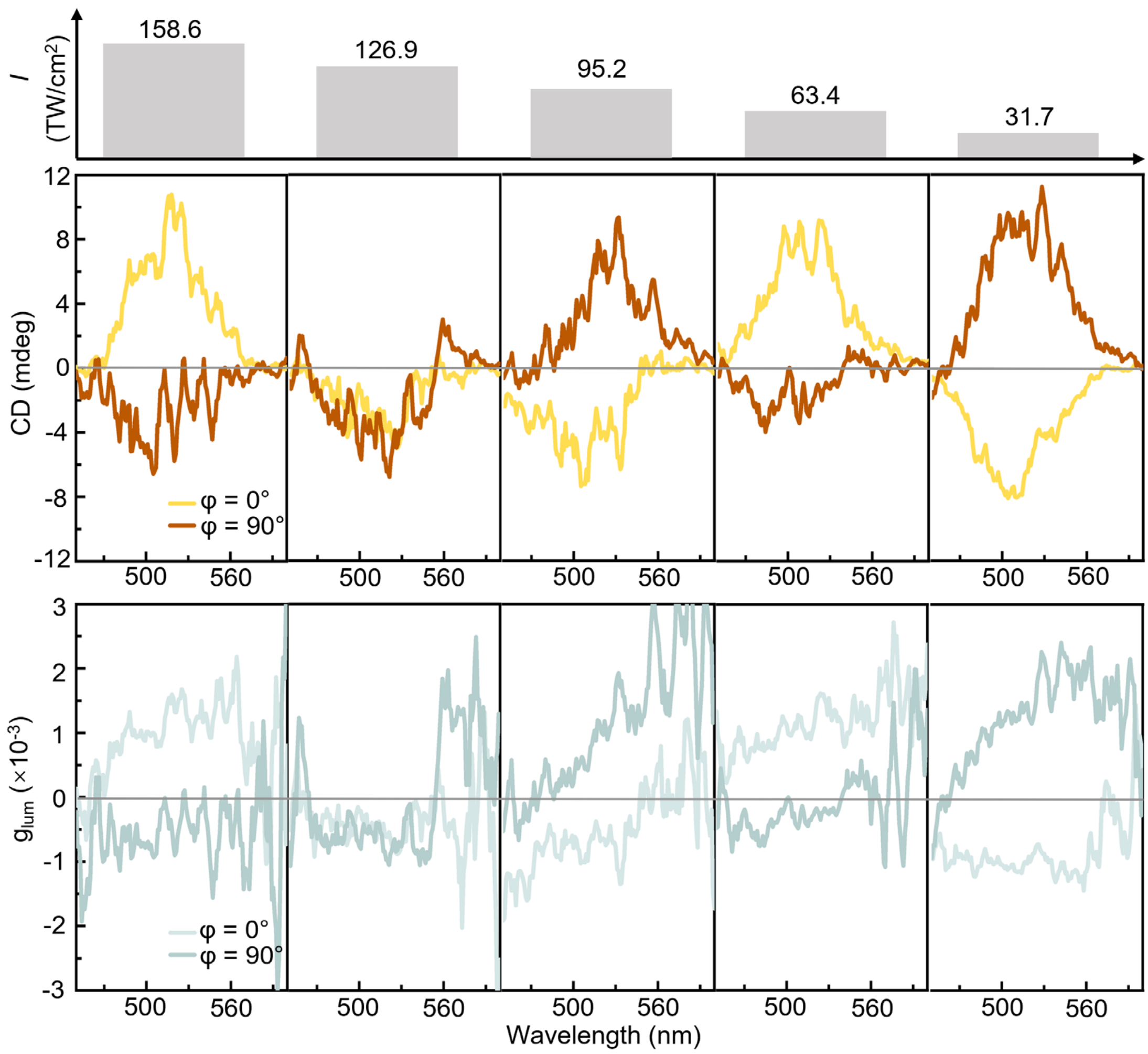

**Figure S8.** CPL responses of laser-written samples without subsequent heat treatment. The samples were fabricated at laser power densities of 158.6, 126.9, 95.2, 63.4, and 31.7 TW/cm$^2$ as indicated from left to right. The CPL spectra were measured at sample azimuthal angles of φ = 0° and 90°.

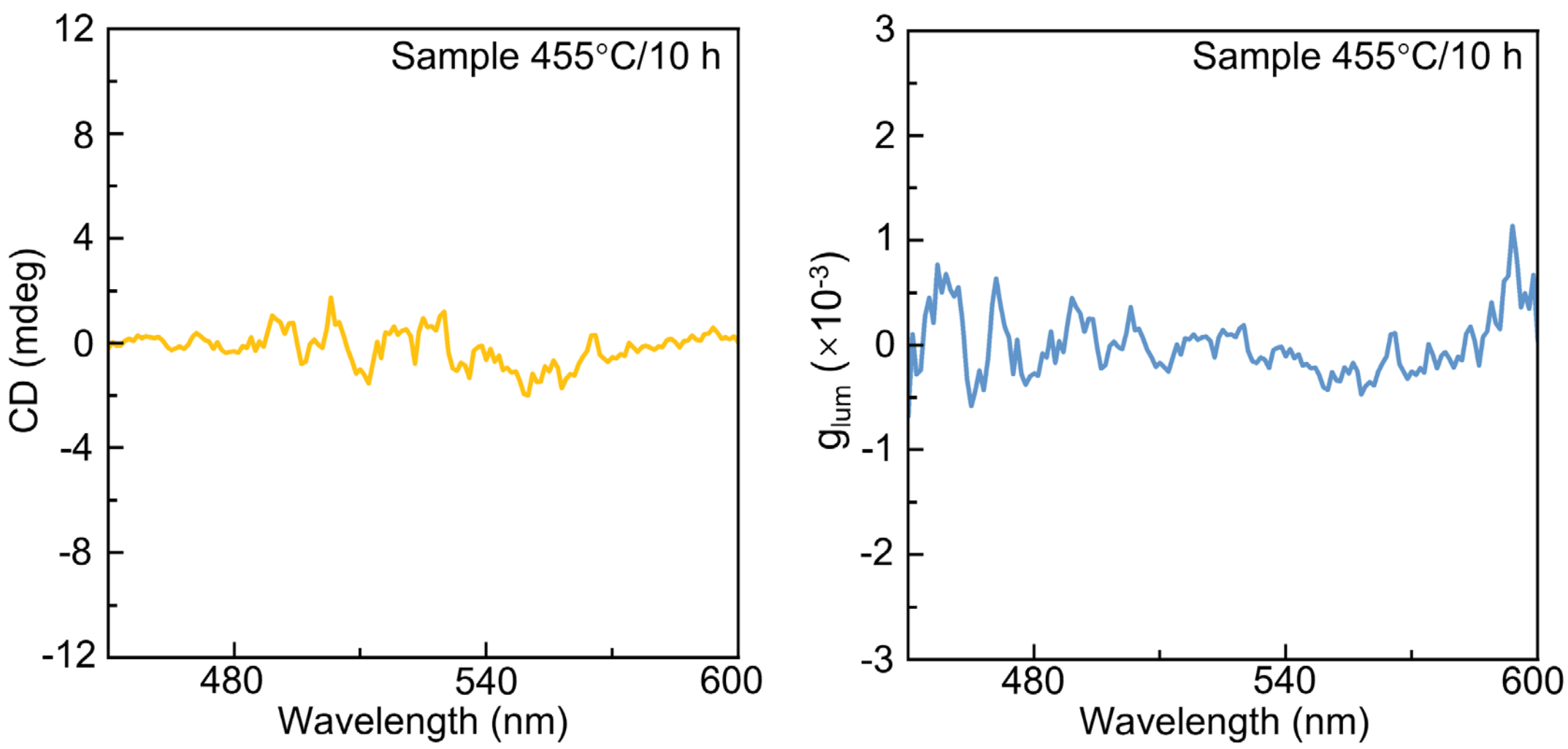

**Figure S9.** CPL response of a uniformly heat-treated sample without prior laser writing. The sample was prepared by thermal treatment at 455 °C for 10 h. The CPL spectrum and the corresponding $g_{lum}$ are shown in the left and right panels, respectively. The measured signals remain close to the experimental baseline over the investigated wavelength range.

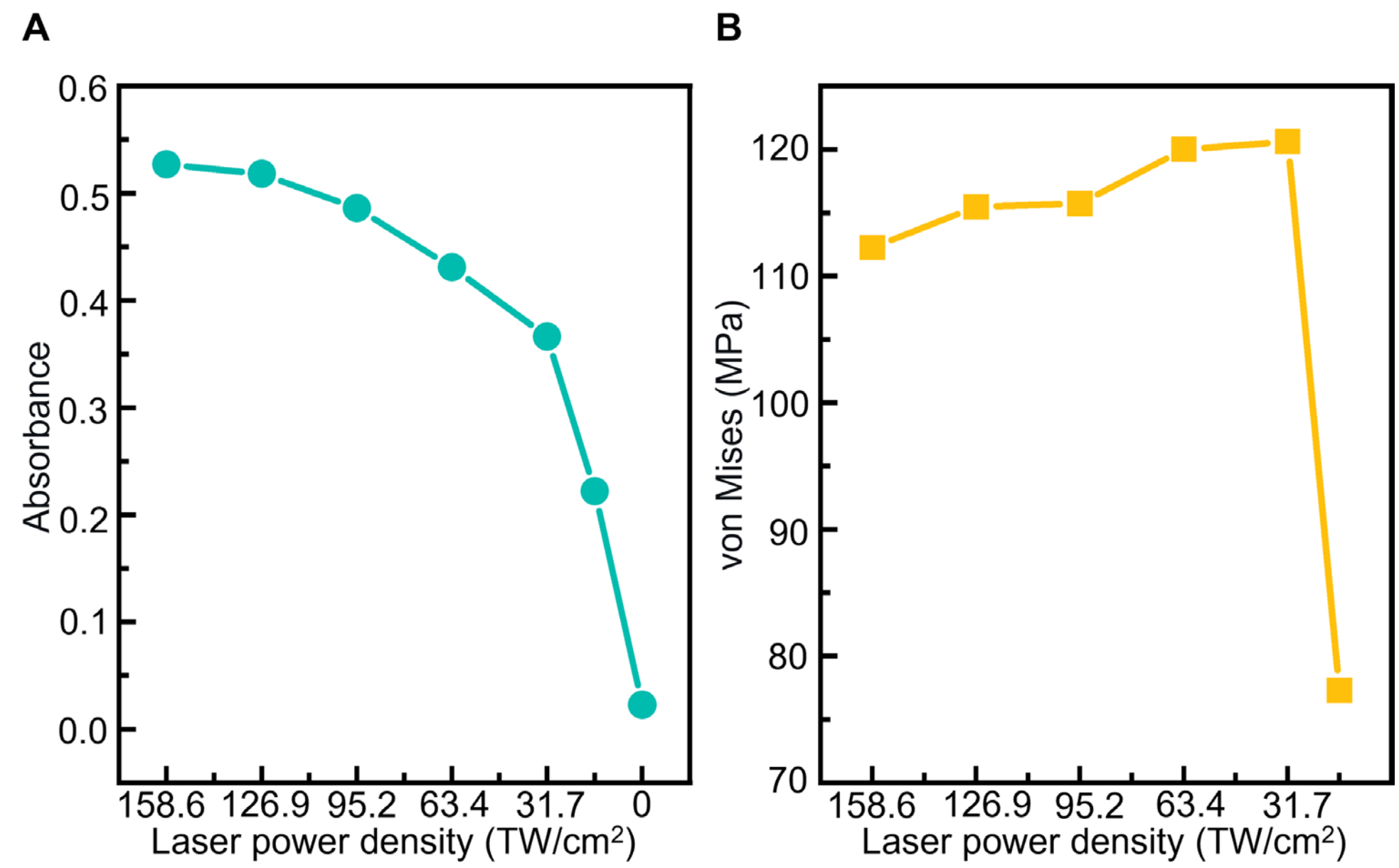

**Figure S10.** Power-dependent absorbance and simulated residual-stress response of the laser-modified glass. (A) Experimentally measured absorbance at the laser-writing wavelength and (B) simulated residual von Mises stress at (x, y, z) = (0, 0, -1 μm) at the end of the modeled post-pulse cooling interval (t = 0.4 ms) as a function of laser power density.

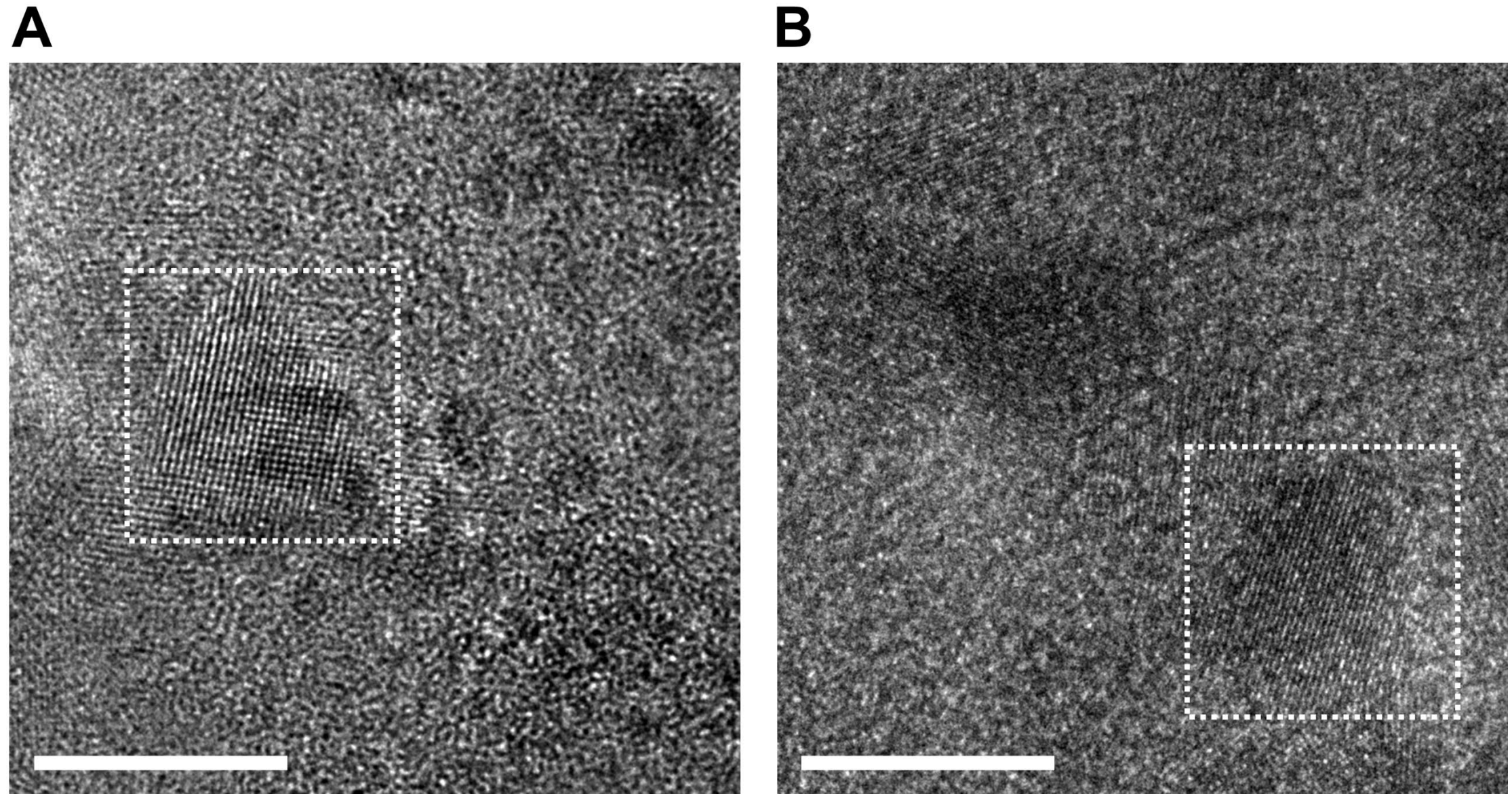

**Figure S11.** HRTEM characterization of PNCs fabricated at a high laser power density of 158.6 $TW/cm^2$. Representative HRTEM images of the laser-written sample before (A) and after (B) subsequent low-temperature heat treatment at 350 °C for 10 h. Scale bars, 10 nm.

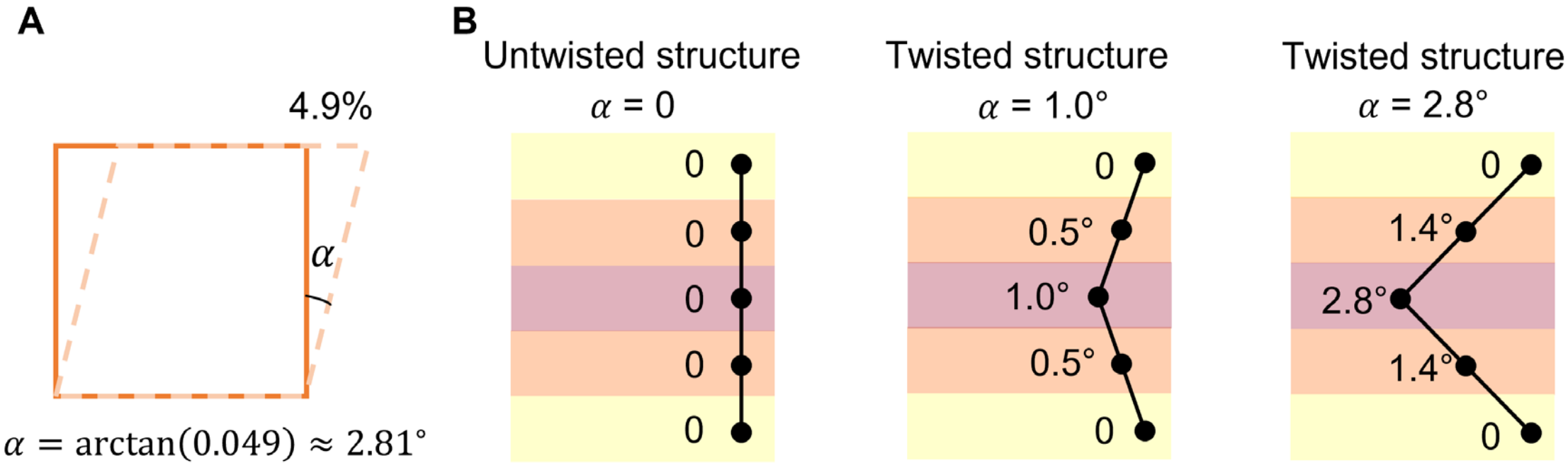

**Figure S12.** Geometric estimation and construction of gradient torsional models for DFT calculations. (A) Geometric estimation of the torsional angle from the apparent interplanar-spacing mismatch. Within the simplified shear-deformation model used here, an apparent interplanar-spacing mismatch of 4.9% corresponds to an estimated torsional angle of $\alpha = \arctan(0.049) \approx 2.8°$. (B) Schematic illustration of the untwisted and gradient torsional models. The untwisted structure corresponds to $\alpha = 0°$. For the twisted structures, the torsional distortion varies symmetrically across the five atomic layers, with the maximum rotation angle at the central layer. The layer-resolved rotation angles are 0°, 0.5°, 1.0°, 0.5°, and 0° for the $\alpha = 1.0°$ model, and 0°, 1.4°, 2.8°, 1.4°, and 0° for the $\alpha = 2.8°$ model.

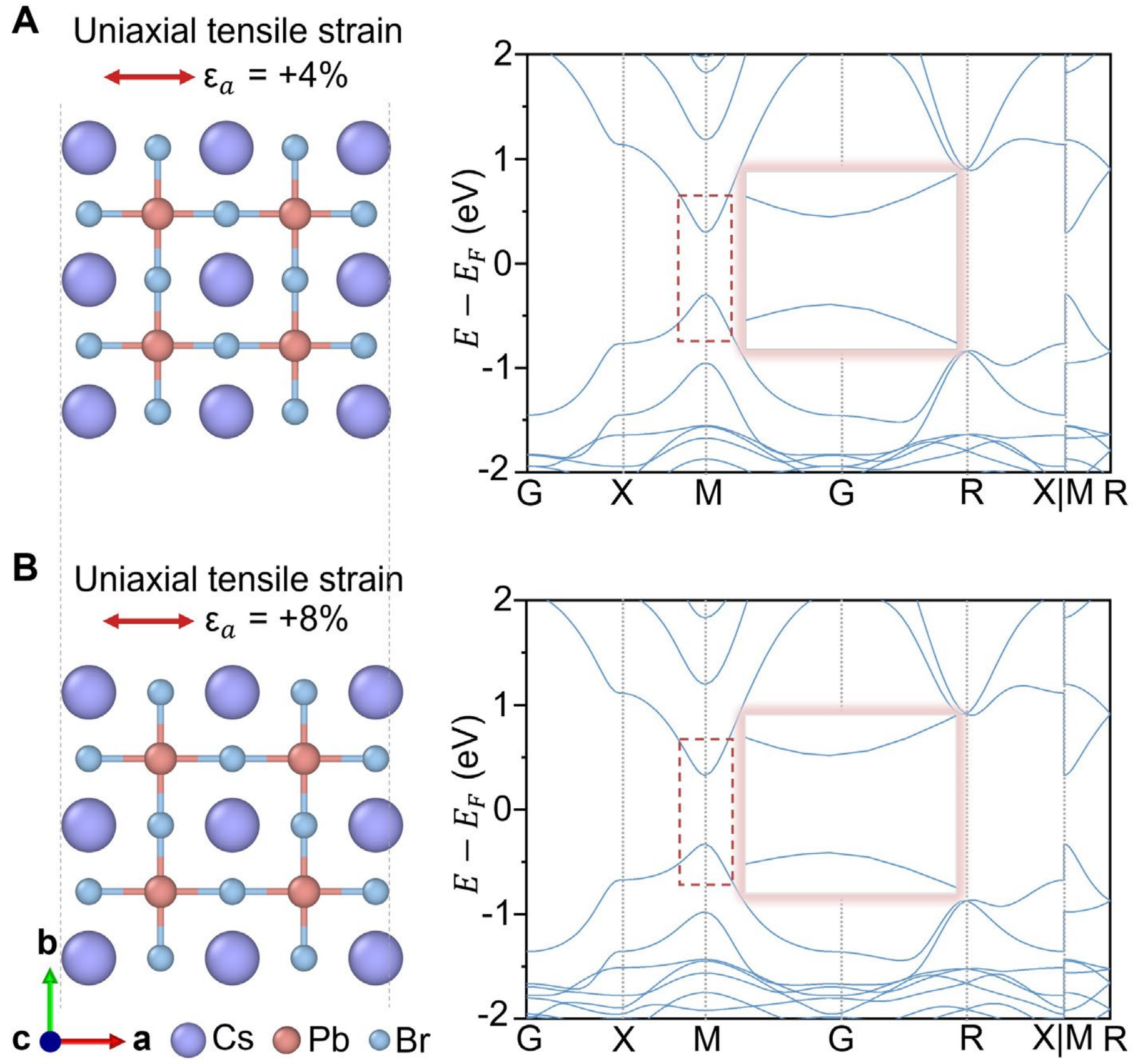

**Figure S13.** DFT-calculated atomic structures and SOC-included electronic band structures of cubic $CsPbBr_3$ under uniaxial tensile strains of 4% (A) and 8% (B) along the *a* axis. The dashed boxes indicate the band-edge regions enlarged in the insets.

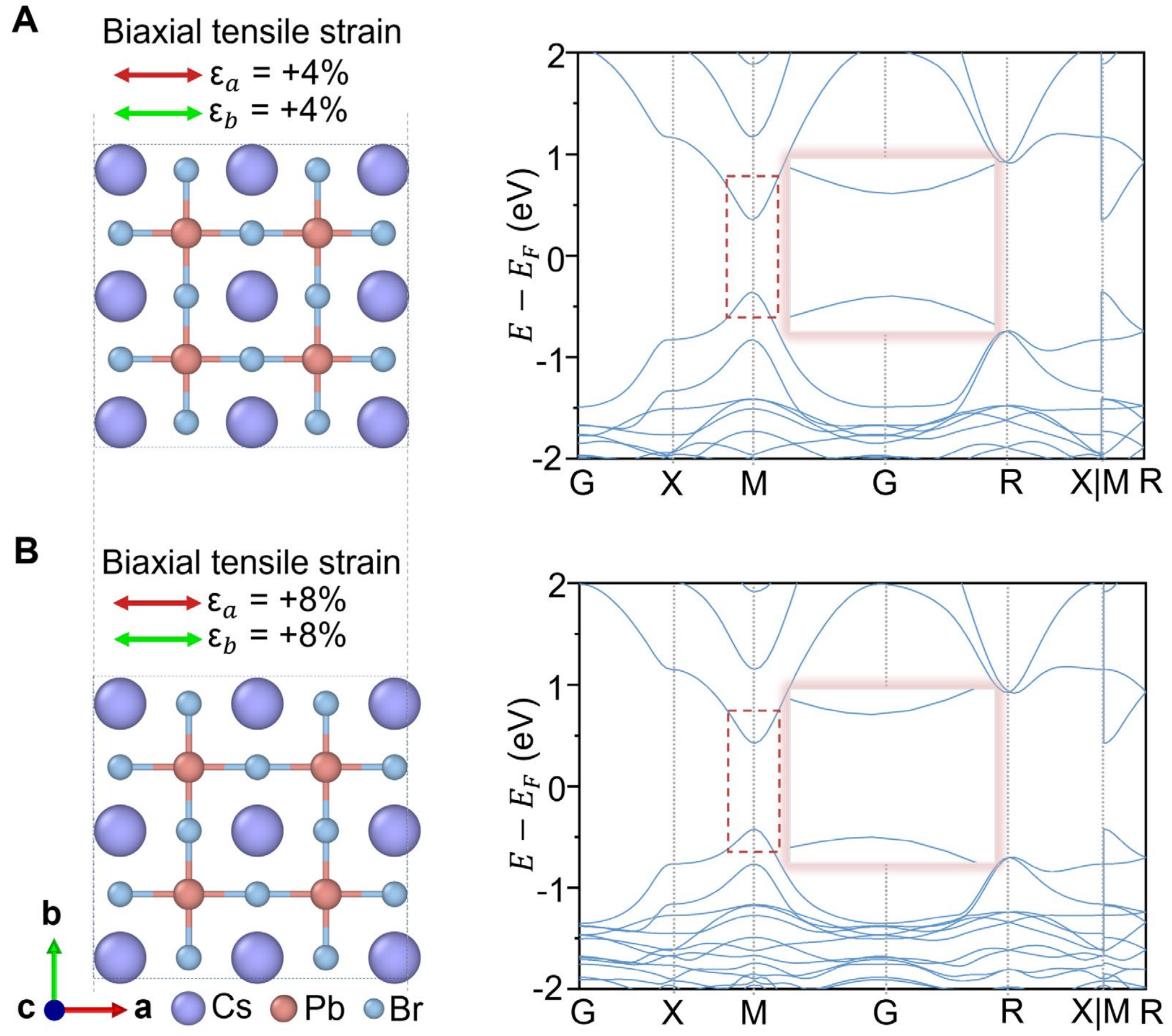

**Figure S14.** DFT-calculated atomic structures and SOC-included electronic band structures of cubic $CsPbBr_3$ under biaxial tensile strains of 4% (A) and 8% (B) along *a* and *c* axes. The dashed boxes indicate the band-edge regions enlarged in the insets.

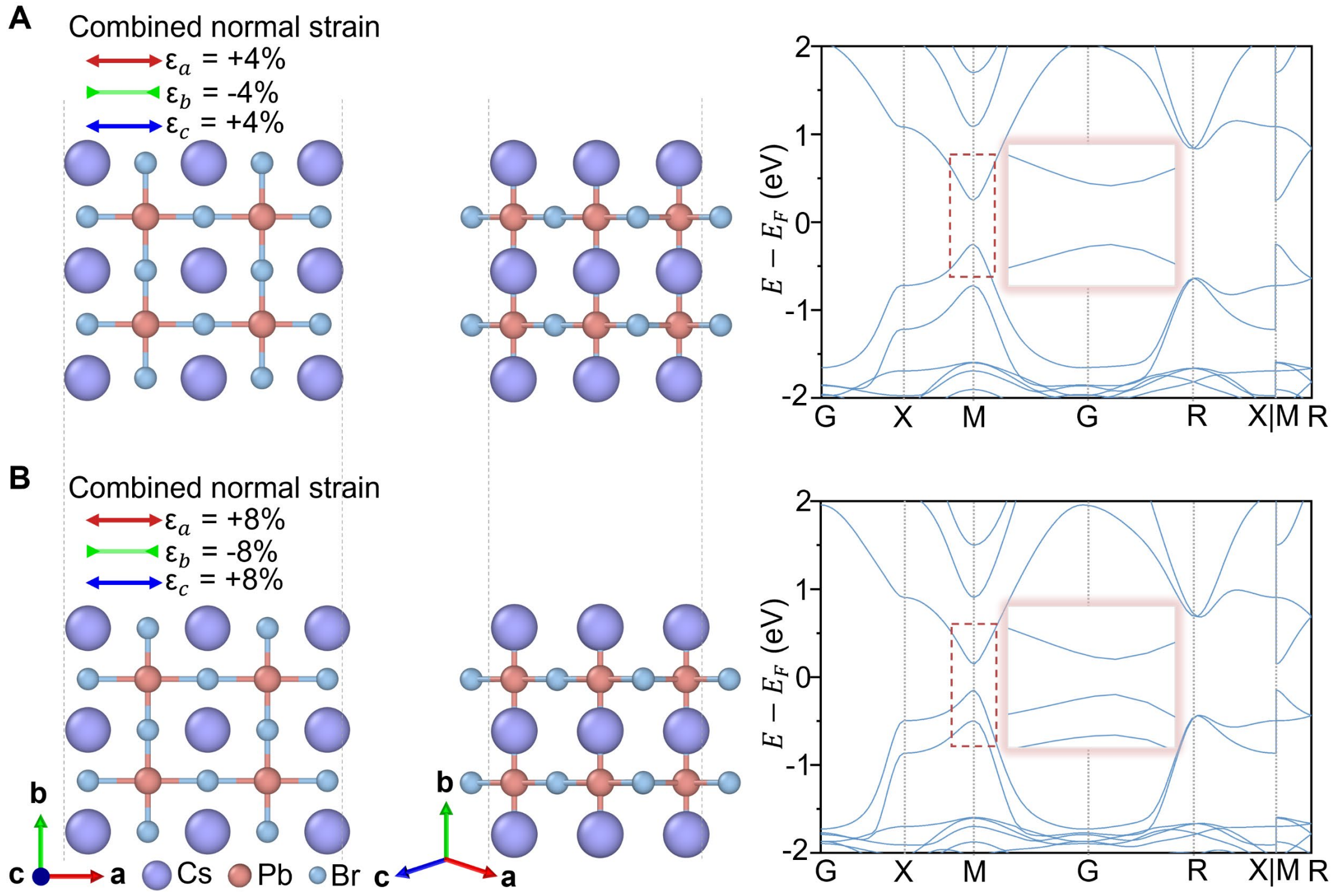

**Figure S15.** DFT-calculated atomic structures and SOC-included electronic band structures of cubic $CsPbBr_3$ under combined normal strains of 4% (A) and 8% (B), with tensile strain applied the along *a* and *c* axes and compressive strain applied along the *b* axis. The dashed boxes indicate the band-edge regions enlarged in the insets.

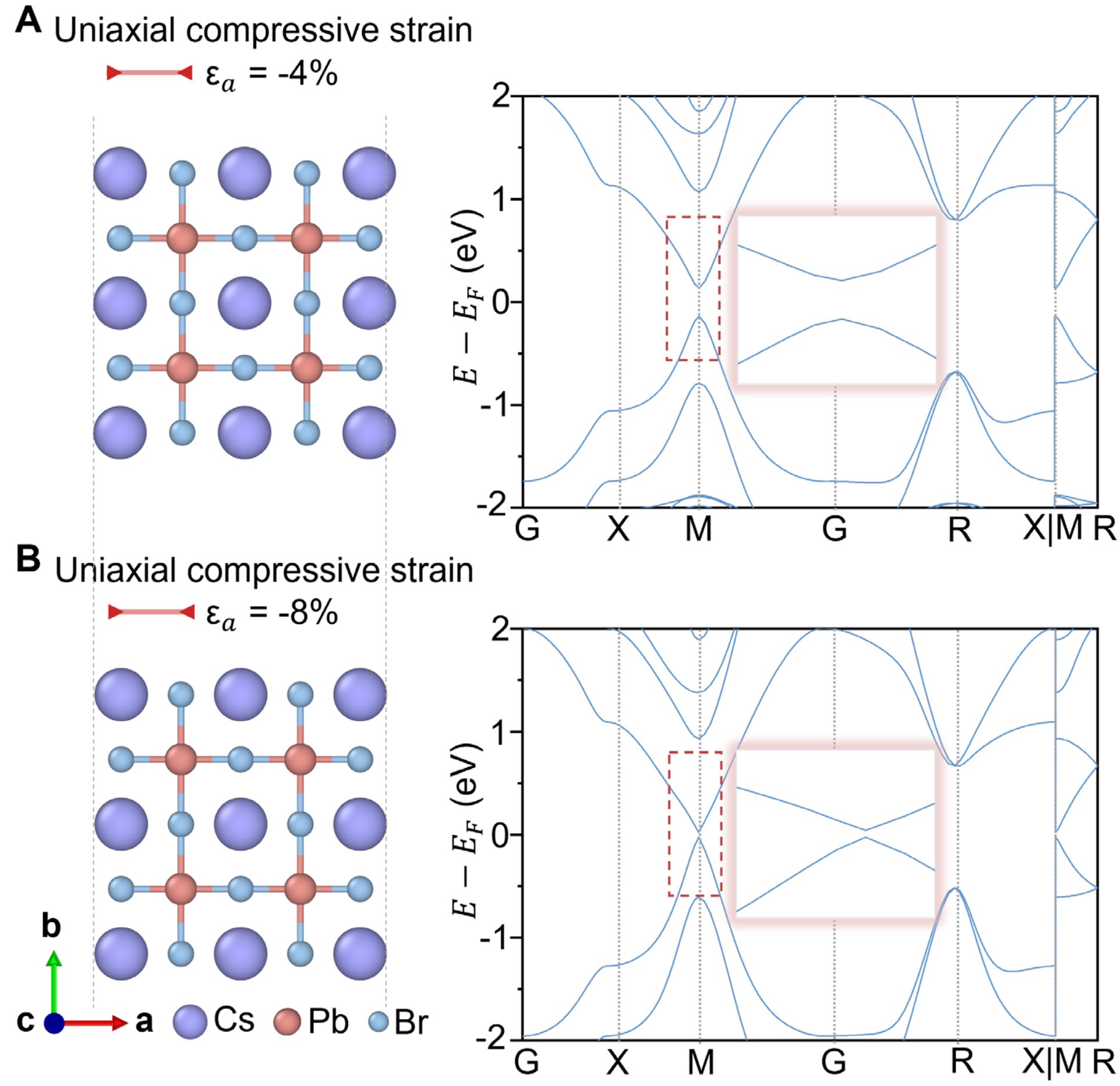

**Figure S16.** DFT-calculated atomic structures and SOC-included electronic band structures of cubic $CsPbBr_3$ under uniaxial compressive strains of 4% (A) and 8% (B) along the *a* axis. The dashed boxes indicate the band-edge regions enlarged in the insets.

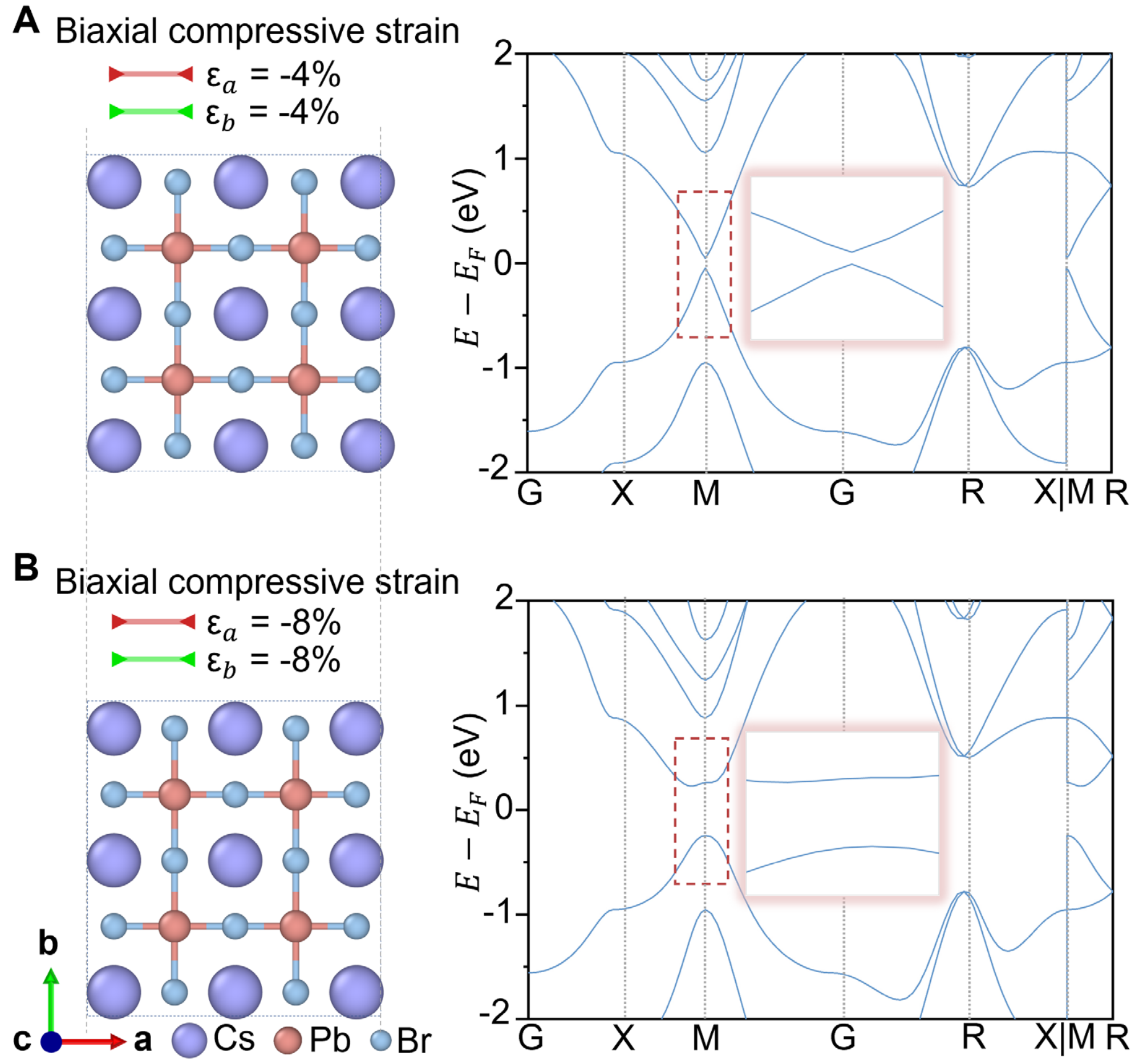

**Figure S17.** DFT-calculated atomic structures and SOC-included electronic band structures of cubic $CsPbBr_3$ under biaxial compressive strains of 4% (A) and 8% (B) along *a* and *b* axes. The dashed boxes indicate the band-edge regions enlarged in the insets.

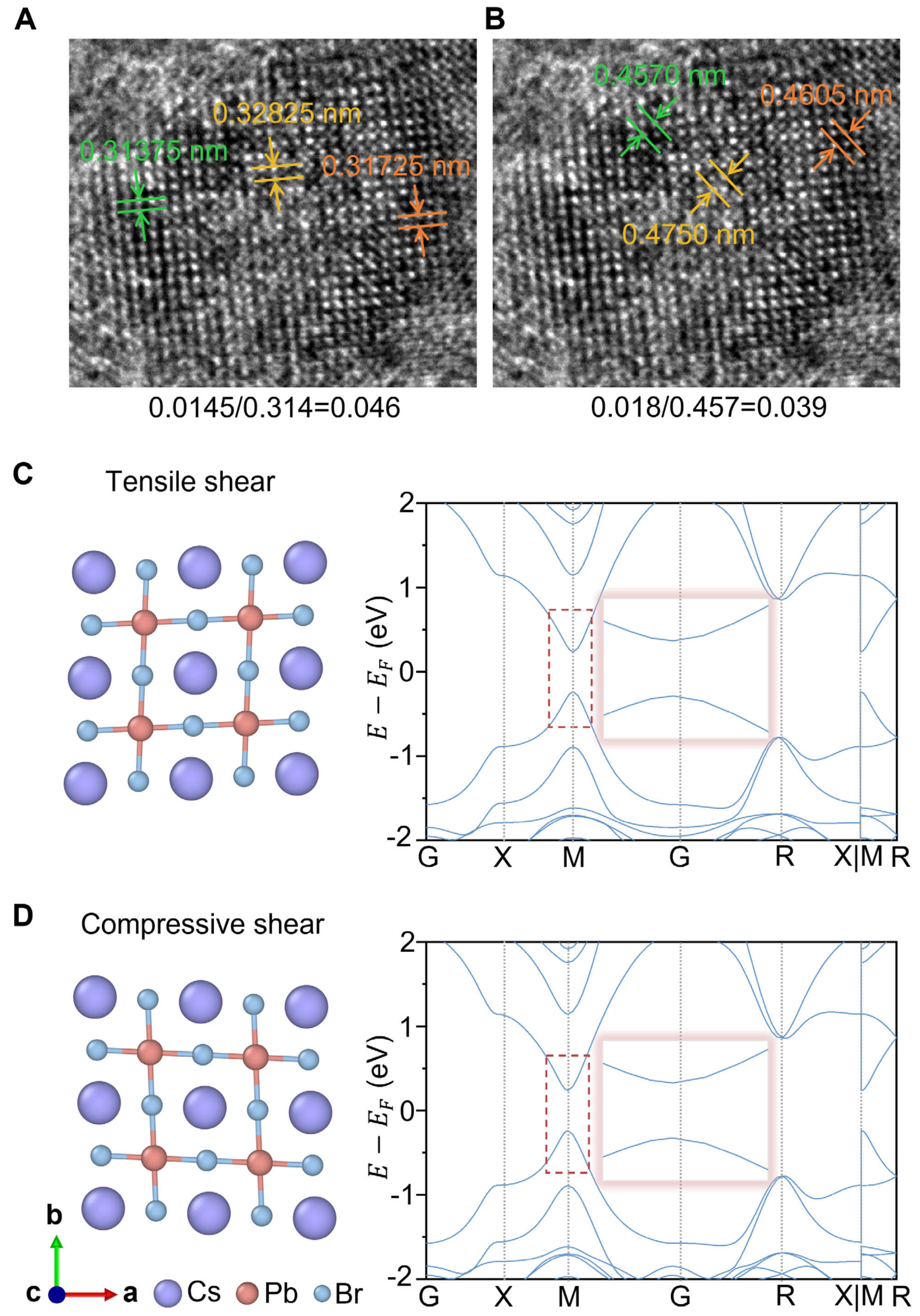

**Figure S18.** Experimental estimation and DFT analysis of shear deformation in $CsPbBr_3$ NCs. (A, B) Representative HRTEM images showing local interplanar-spacing differences corresponding to relative mismatches of approximately 4.6% and 3.9%, respectively. (C, D) DFT-calculated atomic structures and SOC-included electronic band structures of cubic $CsPbBr_3$ under tensile and compressive shear deformation, respectively. The dashed boxes indicate the band-edge regions enlarged in the insets.

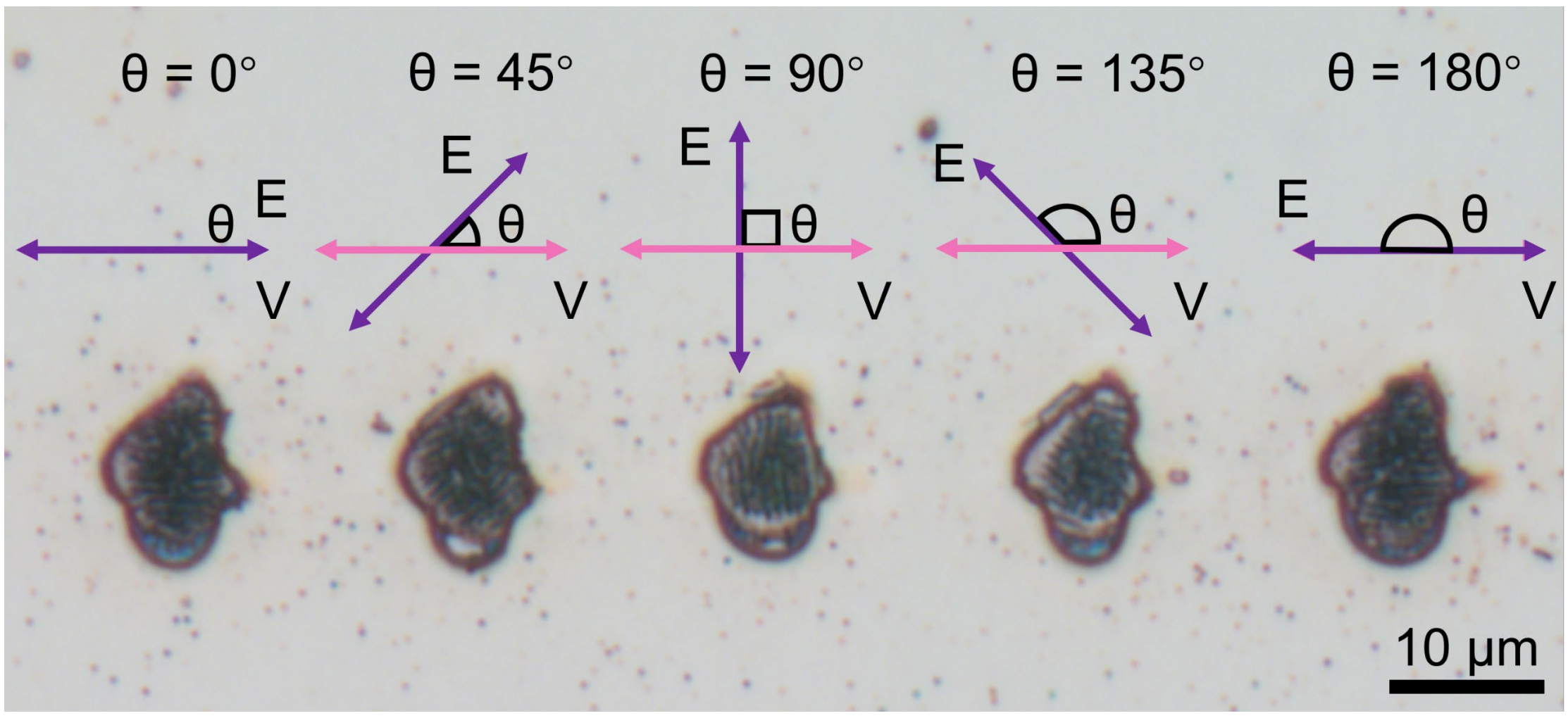

**Figure S19.** Polarization-dependent laser-induced surface morphologies used as a qualitative reference for the stripe orientation in the COMSOL simulations. A silicon-coated titanium film was irradiated with a linearly polarized laser at polarization-writing-axis angles of $\theta$ = 0°, 45°, 90°, 135° and 180°. Here, E denotes the linear-polarization axis, V denotes the laser-writing axis, and $\theta$ is the angle between the two axes. The polarization-dependent stripe orientations observed in this reference experiment were used to define the corresponding orientations of the idealized stripe-modulated volumetric heat sources in the transient thermo-mechanical simulations in Figures 4C and 4D; they were not used as a quantitative representation of the internal energy-deposition profile in the perovskite glass. The laser wavelength, pulse duration, repetition frequency, objective NA, laser power density, and number of pulses delivered per spot were 1030 nm, 5 ps, 1 kHz, 0.3, 0.52 TW/cm$^2$, and 50, respectively. Scale bar, 10 μm.

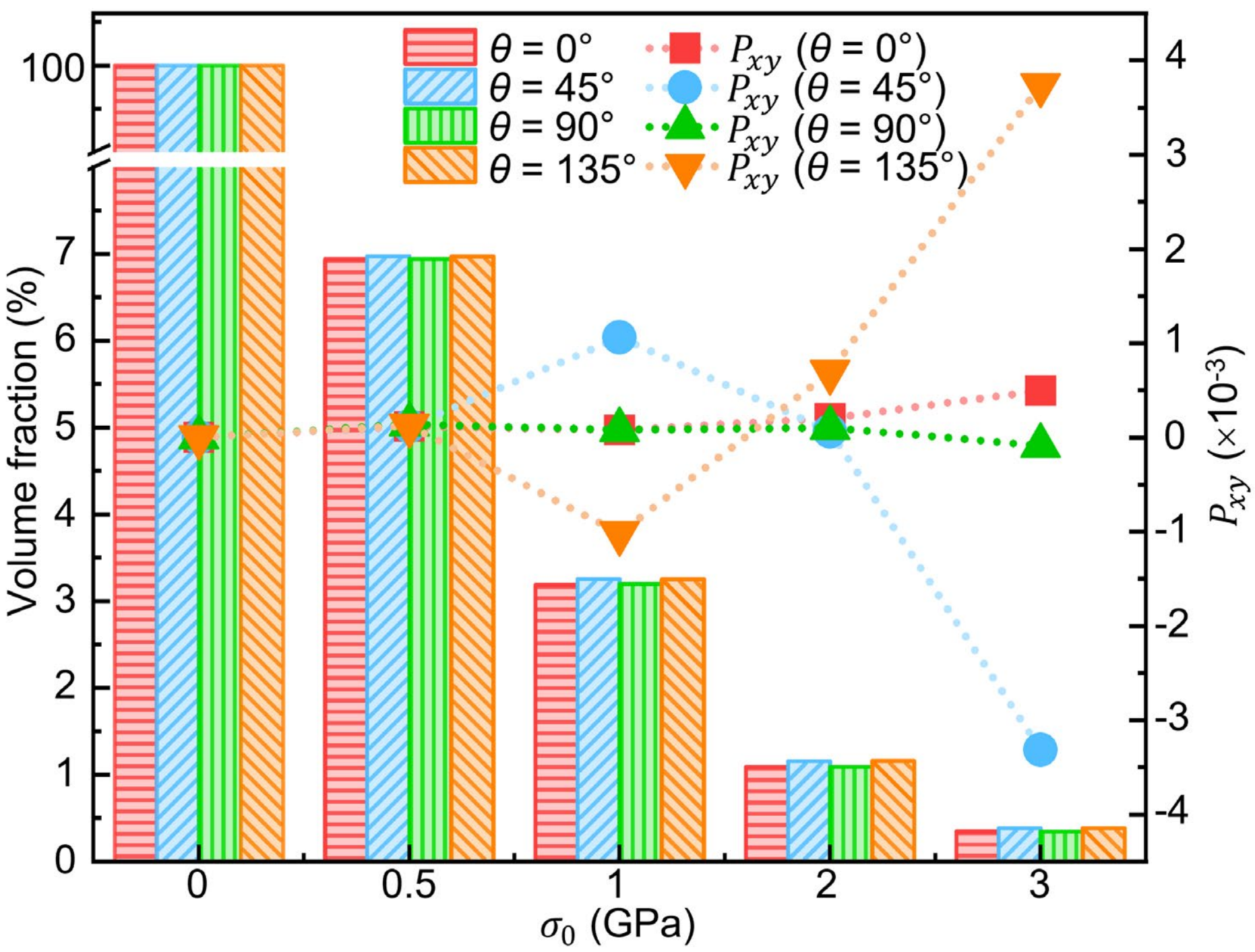

**Figure S20.** Threshold-dependent analysis of the simulated thermally induced $xy$-shear stress. The hatched bars represent the volume fraction of the computational region satisfying $|\sigma_{\mathrm{xy}}| > \sigma_0$, where $\sigma_{\mathrm{xy}}$ is the local $xy$-shear-stress component and $\sigma_0$ is a selected shear-stress magnitude threshold, for polarization-writing-axis angles of $\theta$ = 0°, 45°, 90° and 135°. The corresponding normalized signed shear parameters $P_{xy}$, evaluated over the same threshold-selected regions, are shown as symbols. The dotted lines are guides to the eye.

**Table S1.** Stokes parameters and degrees of polarization of the laser-written sample. The sample was excited using a 425 nm continuous-wave laser with a power of 70 μW through a 50× objective. The sample was fabricated at a laser power density of 63.4 TW/cm$^2$, followed by low-temperature heat treatment at 350 °C for 10 h.

| Parameter | Value |
|---|---|
| $s_1 = \frac{S_1}{S_0}$ | 0.1949 |
| $s_2 = \frac{S_2}{S_0}$ | 0.0020 |
| $s_3 = \frac{S_3}{S_0}$ | -0.0080 |
| $DOLP = \frac{\sqrt{S_1^2 + S_2^2}}{S_0}$ | 0.1949 |
| $DOCP = \frac{\lvert S_3 \rvert}{S_0}$ | 0.0080 |
| $DOP = \frac{\sqrt{S_1^2 + S_2^2 + S_3^2}}{S_0}$ | 0.1951 |

**Table S2.** Thermal and viscoelastic parameters used in the power-dependent residual-stress simulation in Note S1.

| Parameter | Value | Unit |
|---|---|---|
| Density | 3800 | $kg/m^3$ |
| Thermal conductivity | 1.1 | W/(m·K) |
| Specific heat capacity | 650 | J/(kg·K) |
| Thermal expansion coefficient | $6 \times 10^{-6}$ | 1/K |
| Reference temperature | 293.15 | K |
| Pulse period | $10^{-4}$ | s |
| Bulk modulus | $4.286 \times 10^{10}$ | Pa |
| Long-term shear modulus | $2.0 \times 10^{7}$ | Pa |

**Table S3.** Generalized Maxwell model parameters. The generalized Maxwell model consisted of six relaxation branches with the shear moduli Gi and relaxation times τi listed below. These effective parameters were used to reproduce the qualitative stress-relaxation behavior of the laser-modified glass.

| $i$ | $G_i$ (Pa) | $\tau_i$ (s) |
|---|---|---|
| 1 | $6.0 \times 10^{9}$ | $1.0 \times 10^{-5}$ |
| 2 | $6.0 \times 10^{9}$ | $1.0 \times 10^{-3}$ |
| 3 | $6.0 \times 10^{9}$ | $1.0 \times 10^{-1}$ |
| 4 | $5.0 \times 10^{9}$ | 1 |
| 5 | $4.0 \times 10^{9}$ | 10 |
| 6 | $2.5 \times 10^{9}$ | $1.0 \times 10^{4}$ |

**Table S4.** Structural distortion parameters of cubic $CsPbBr_3$ for the untwisted structure and twisted structures with imposed maximum twisting angles of 1.0° and 2.8°.

| Maximum twisting angle | $D$ | $\sigma^2$ | $\beta$ | $\Delta\beta$ |
|---|---|---|---|---|
| 0.0° (No twist) | $3.68 \times 10^{-7}$ | 0.000087 | 180.0° | 0.0° |
| 1.0° | 0.000038 | 0.363870 | 178.6°, 179.0° | 0.4° |
| 2.8° | 0.000244 | 2.849912 | 176.0°, 177.2° | 1.2° |

**Table S5.** Laser-writing parameters for the multichannel optical encoding pattern, at power densities of 126.9 $TW/cm^2$ and 63.4 $TW/cm^2$.

| Area | Wavelength (nm) | Pulse width (ps) | Repetition rate (kHz) | Objective lens | Laser power density ($TW/cm^2$) | $\boldsymbol{\theta}$ (°) |
|---|---|---|---|---|---|---|
| 1 | 1030 | 5 | 10 | 10× | 63.4 | 135 |
| 2 | 1030 | 5 | 10 | 10× | 126.9 | 45 |
| 3 | 1030 | 5 | 10 | 10× | 63.4 | 135 |
| 4 | 1030 | 5 | 10 | 10× | 126.9 | 45 |
| 5 | 1030 | 5 | 10 | 10× | 63.4 | 45 |
| 6 | 1030 | 5 | 10 | 10× | 126.9 | 45 |
| 7 | 1030 | 5 | 10 | 10× | 63.4 | 135 |
| 8 | 1030 | 5 | 10 | 10× | 126.9 | 45 |
| 9 | 1030 | 5 | 10 | 10× | 63.4 | 135 |